\documentclass[iop]{emulateapj} 
\usepackage{apjfonts} 
\usepackage{epstopdf}
\usepackage{graphicx}
\def\alwaysmath#1{\ifmmode{#1}\else{$#1$}\fi} 

\usepackage[pdfpagemode={UseOutlines},bookmarks=true,bookmarksopen=true,
   bookmarksopenlevel=0,bookmarksnumbered=true,hypertexnames=false,
   colorlinks,linkcolor={blue},citecolor={blue},urlcolor={red},
   pdfstartview={FitV},unicode,breaklinks=true]{hyperref}
 
\lefthead{Hasselquist et al.} 
\righthead{Identifying Sgr Stream Stars}
\newcommand{\Beta}{B}

\begin{document} 
\DeclareGraphicsExtensions{.ps,.pdf,.png,.jpg,.eps}

\title{Identifying Sagittarius Stream Stars By Their APOGEE Chemical Abundance Signatures} 
 
\author{Sten Hasselquist\altaffilmark{1,2,$\dagger$}, 
Jeffrey L. Carlin\altaffilmark{3},
Jon A. Holtzman\altaffilmark{1},
Matthew Shetrone\altaffilmark{4},
Christian R. Hayes\altaffilmark{7},
Katia Cunha\altaffilmark{12},
Verne Smith\altaffilmark{5},
Rachael L. Beaton\altaffilmark{6},
Jennifer Sobeck\altaffilmark{13},  
Carlos Allende Prieto\altaffilmark{10,11},
Steven R. Majewski\altaffilmark{7},
Borja Anguiano\altaffilmark{7},
Dmitry Bizyaev\altaffilmark{15,16},
D. A. Garc{\'{\i}}a-Hern{\'a}ndez\altaffilmark{10,11},
Richard R. Lane\altaffilmark{8,9},
Kaike Pan\altaffilmark{15},
David L. Nidever\altaffilmark{14,5},
Jos\'e. G. Fern\'andez-Trincado\altaffilmark{17,18,19},
John C. Wilson\altaffilmark{7},
Olga Zamora\altaffilmark{10,11}
}

\altaffiltext{1}{New Mexico State University, Las Cruces, NM 88003, USA (sten@nmsu.edu)}

\altaffiltext{2}{Department of Physics \& Astronomy, University of Utah, Salt Lake City, UT, 84112, USA (stenhasselquist@astro.utah.edu)}

\altaffiltext{$\dagger$}{NSF Astronomy and Astrophysics Postdoctoral Fellow}

\altaffiltext{3}{LSST, Tucson, AZ 85719, USA (jcarlin@lsst.org)}

 \altaffiltext{4}{University of Texas at Austin, McDonald Observatory, Fort Davis, TX 79734, USA
 (shetrone@astro.as.utexas.edu)}

 \altaffiltext{5}{National Optical Astronomy Observatories, Tucson, AZ 85719, USA (vsmith, dnidever@email.noao.edu)}

\altaffiltext{6}{Department of Astrophysical Sciences, Princeton University, 4 Ivy Lane, Princeton, NJ 08544, USA (rachael.l.beaton@gmail.com)}

 \altaffiltext{7}{Dept. of Astronomy, University of Virginia, 
 Charlottesville, VA 22904-4325, USA (crh7gs, srm4n, ba7t, jcw6z@virginia.edu)}
 
 \altaffiltext{8}{Instituto de Astrof{\'{\i}}sica, Pontificia Universidad Cat\'olica de Chile, Av. Vicuna Mackenna 4860, 782-0436 Macul, Santiago, Chile (rlane@astro.puc.cl)}
 
 \altaffiltext{9}{Millennium Institute of Astrophysics, Av. Vicu\~na Mackenna 4860, 782-0436 Macul, Santiago, Chile (rlane@astro.puc.cl)}

\altaffiltext{10}{Instituto de Astrof{\'{\i}}sica de Canarias, E-38205 La Laguna, Tenerife, Spain (callende, agarcia, ozamora@iac.es)}

\altaffiltext{11}{Universidad de La Laguna (ULL), Departamento de Astrof{\'{\i}}sica, E-38206 La Laguna, Tenerife, Spain (callende, agarcia, ozamora@iac.es)}

  \altaffiltext{12}{Observat\'orio Nacional/MCTI, Rua Gen. Jos\'e Cristino, 77, 20921-400, Rio de Janeiro, Brazil (cunha@email.noao.edu}
  
  \altaffiltext{13}{Department of Astronomy, Box 351580, University of Washington, Seattle, WA 98195, USA (jsobeck@uw.edu)}
  
  \altaffiltext{14}{Department of Physics, Montana State University, P.O. Box 173840, Bozeman, MT 59717-3840 (dnidever@email.noao.edu)}

\altaffiltext{15}{Apache Point Observatory and New Mexico State University, P.O. Box 59, Sunspot, NM, 88349-0059, USA}

\altaffiltext{16}{Sternberg Astronomical Institute, Moscow State University, Moscow, Russia}

\altaffiltext{17}{Departamento de Astronom\'ia, Casilla 160-C, Universidad de Concepci\'on, Concepci\'on, Chile  (jfernandezt@astro-udec.cl)}

\altaffiltext{18}{Instituto de Astronom\'ia y Ciencias Planetarias, Universidad de Atacama, Copayapu 485, Copiap\'o, Chile.}

\altaffiltext{19}{Institut Utinam, CNRS UMR 6213, Universit\'e Bourgogne-Franche-Comt\'e, OSU THETA Franche-Comt\'e, Observatoire de Besan\c{c}on, \\ BP 1615, 25010 Besan\c{c}on Cedex, France.}

\begin{abstract} 
	
The SDSS-IV Apache Point Observatory Galactic Evolution Experiment (APOGEE) survey provides precise chemical abundances of 18 chemical elements for $\sim$ 176,000 red giant stars distributed over much of the Milky Way Galaxy (MW), and includes observations of the core of the Sagittarius dwarf spheroidal galaxy (Sgr). The APOGEE chemical abundance patterns of Sgr have revealed that it is chemically distinct from the MW in most chemical elements. We employ a \emph{k}-means clustering algorithm to 6-dimensional chemical space defined by [(C+N)/Fe], [O/Fe], [Mg/Fe], [Al/Fe], [Mn/Fe], and [Ni/Fe] to identify 62 MW stars in the APOGEE sample that have Sgr-like chemical abundances. Of the 62 stars, 35 have \emph{Gaia} kinematics and positions consistent with those predicted by \emph{N}-body simulations of the Sgr stream, and are likely stars that have been stripped from Sgr during the last two pericenter passages ($<$ 2 Gyr ago). Another 20 of the 62 stars exhibit chemical abundances indistinguishable from the Sgr stream stars, but are on highly eccentric orbits with median $r_{\rm apo} \sim $ 25 kpc. These stars are likely the ``accreted'' halo population thought to be the result of a separate merger with the MW 8-11 Gyr ago. We also find one hypervelocity star candidate. We conclude that Sgr was enriched to [Fe/H] $\sim$ -0.2 before its most recent pericenter passage. If the ``accreted halo'' population is from one major accretion event, then this progenitor galaxy was enriched to at least [Fe/H] $\sim$ -0.6, and had a similar star formation history to Sgr before merging.  
\end{abstract}

 
\section{Introduction}

The stellar halo of the Milky Way (MW) contains only a small fraction of the total stellar mass of our Galaxy, but serves as an important laboratory to study the accretion history of the MW. N-body simulations suggest that galaxies like our own are formed, at least in part, through hierarchical buildup from smaller stellar systems (e.g., \citealt{Klypin1999}). Therefore, accreted systems should be able to be observed in various states of dissolution as they are destroyed upon entering the MW potential well. In fact, deep photometric surveys have revealed that the distribution of stars that make up the Milky Way stellar halo actually contains spatially extended and relatively massive stellar substructure, typically in the form of stellar streams (see, e.g., \citealt{Belokurov2006,Schlaufman2011,Schlaufman2012}, or a review of known systems in \citealt{Grillmair2016}). Some of these streams can be linked to obvious progenitors, such as the Sagittarius dwarf spheroidal galaxy (Sgr, \citealt{Ibata1994,Majewski2003}), while other streams appear to be ``orphans'', with the progenitor system unknown (e.g., \citealt{Belokurov2007}).

While strong evidence for substructure in the MW stellar halo has been observed, it is not clear as to how much of the halo was formed in situ versus was accreted, or even what fraction of the halo consists of disk stars that have been kicked out into highly eccentric orbits (see, e.g., \citealt{Sheffield2012}). Simulations have revealed that the stellar halo was likely formed through some combination of in situ star formation and accretion, with the inner halo (R $\lesssim$ 20 kpc) having a larger fraction of stars formed in situ than the outer halo (e.g., \citealt{Bell2008,Zolotov2009,Font2011}). Such features have been found in observations of the MW halo (e.g., \citealt{Carollo2007,Beers2012}). Other observations have revealed that there is a break in the density profile of the halo that is coincident with a change in slope of the metallicity gradient, which shifts from nearly flat to negative at R $\gtrsim$ 20 kpc (e.g., \citealt{AllendePrieto2014,Xue2015,Fernandez-Alvar2015}), lending support to these simulations. Simulations of accretion scenarios find that the substructure should remain spatially and kinematically coherent throughout much of the lifetime of the MW (e.g., \citealt{Johnston1996}), so it is likely more substructure will be identified as photometric and spectroscopic surveys of the entire sky continue (e.g., Pan-STARRS, \emph{Gaia}, and LSST).  In fact, recent works utilizing \emph{Gaia} data suggest that the inner Galactic halo is dominated by debris from a massive, LMC-sized merger some 8-11 Gyr ago (e.g., \citealt{Belokurov2018,Deason2018,Koppelman2018,Helmi2018}), that may have also deposited several globular clusters into the MW potential \citep{Myeong2018}. 

It is also possible to find halo substructure in chemical abundance space rather than position-velocity space. In principle, stars should contain a unique fingerprint in their detailed chemical abundance patterns that link them to the same molecular cloud of their birth, or at least, a molecular cloud that was polluted by metals in similar ways. If true, then this suggests that measuring the detailed chemical abundances of stars can reveal which stars were born from the same molecular cloud. This technique, known as ``chemical tagging'' \citep{Freeman&Bland-Hawthorn2002}, requires large spectroscopic surveys that can observe hundreds of thousands, or even millions of stars, and that are capable of providing precise chemical abundances. Fortunately, surveys like these have recently been completed or are ongoing (e.g., RAVE, SDSS, LAMOST, GALAH, etc.), and this technique of chemical tagging can be explored (e.g., \citealt{Mitschang2014}).

Here we focus on the Apache Point Observatory Galactic Evolution Experiment (APOGEE, \citealt{Majewski2017}). APOGEE has observed 176,000 red giant stars across the MW from the northern hemisphere, and as part of APOGEE-2, has recently begun observations in the Southern Hemisphere. By 2020, APOGEE will have detailed chemical abundances for $\sim$ 500,000 stars sampling the entire MW, opening up a vast Galactic volume to chemical tagging experiments. Because APOGEE observes in the near infrared ($H$-band, 1.5-1.7 $\mu$m) and targets red giant stars that are brighter in the NIR than optical, it is able to see stars as far as $\sim$ 10 kpc in the dusty plane of the MW and over 60 kpc out into the halo. This makes it an ideal instrument to study the accretion of the MW halo, especially if accreted populations do dominate at R $>$ 20 kpc.

Chemical tagging techniques have already been explored using the APOGEE data set. \citet{Hogg2016} applied a \emph{k}-means clustering algorithm to show that abundances provided by the data-driven algorithm ``The Cannon'' \citep{Ness2015} are precise enough to recover groups of stars that can be traced to Sgr and some globular clusters. However, \citet{Ness2017} found that chemical tagging, in its purest form of linking stars to the same birth molecular cloud, is not feasible with the APOGEE data. They found that with the  APOGEE abundances from The Cannon, where chemical elements are determined down to the 0.04 dex or less level, chemically similar pairs of stars are more likely to be ``doppelgangers'' (stars that just happen to exhibit the same abundance patterns) rather than siblings actually born from the same molecular cloud. Despite this realization, so called ``weak'' chemical tagging (chemically tagging stars born in the same type of stellar system rather than the same unique stellar system) has been used in the APOGEE dataset to discover field stars that exhibit abundance patterns similar to second-generation globular clusters (e.g., \citealt{Fernandez-Trincado2016,Fernandez-Trincado2017,Schiavon2017}), providing constraints on what fraction of the MW bulge was formed from accreted globular clusters. \citet{Hayes2018a} also found that APOGEE stars with [Fe/H] $<$ $-$0.9 can be split into a high-magnesium population and a low-magnesium population, the latter of which is likely an accreted population based on its chemical abundance patterns, which are similar to those of dwarf galaxies. 

In this work, we explore whether or not APOGEE has the ability to chemically tag stars that have formed in dwarf galaxies by performing a chemical tagging analysis targeted at the Sgr system. Observations as well as models of the Sgr system suggest that Sgr tidal debris covers a large fraction of the MW (e.g., \citealt{Ibata1995,Majewski2003,Belokurov2006,Law&Majewski2010,Tepper-Garcia2018}). The \citet{Law&Majewski2010} model predicts that many Sgr stars should be coincident on the sky with fields observed by APOGEE. It has been shown that the chemical abundance patterns of Sgr are quite unique. The metal-rich Sgr stars ([Fe/H] $>$ $-$0.8) exhibit deficiencies in all chemical elements (expressed as [X/Fe]) relative to the MW (see, e.g., \citealt{McWilliam2013,Hasselquist2017}). Recent work confirms these abundance patterns extend to the Sgr tidal streams as well \citep{Carlin2018}.

The paper is organized as follows. In \S \ref{obs} we explain the sample of APOGEE data in which we look for Sgr stream members. The process for identifying Sgr stream candidates in chemical abundance space is explained in \S \ref{clust}. Results are presented in \S \ref{res}, where we present the discovery of 68 potential Sgr stream members (62 of which have \emph{Gaia} proper motion measurements). Implications for chemical tagging methods as well as constraints on models for the Sgr system are described in \S \ref{sec:disc}.

\section{Observations and Data Reduction}
\label{obs}

\subsection{APOGEE}

The APOGEE survey was part of Sloan Digital Sky Survey III \citep{Eisenstein2011}, and observed 146,000 stars in the Milky Way galaxy \citep{Majewski2017} from 2011-2014. APOGEE-2 began observations in 2014 as part of the Sloan Digital Sky Survey IV \citep{Blanton2017}, and the latest data release, DR14, contains APOGEE observations of an additional $\sim$ 100,000 stars in the Northern Hemisphere \citep{dr142018}. The APOGEE instrument is a high-resolution (R $\sim$ 22,500) near-infrared (1.51-1.70 $\mu$m) spectrograph described in detail in Wilson et al. (in prep). For the main survey the instrument was connected to the Sloan 2.5m telescope \citep{Gunn2006}. Targeting strategies for APOGEE and APOGEE-2 are described in \citet{Zasowski2013} and \citet{Zasowski2017}, respectively.

The APOGEE data are reduced through methods described by \citet{Nidever2015}, and stellar parameters/chemical abundances are extracted using the APOGEE Stellar Parameters and Chemical Abundances Pipeline (ASPCAP, \citealt{Garcia2016}). ASPCAP interpolates in a grid of synthetic spectra \citep{Zamora2015} generated from the APOGEE $H$-band line list \citep{Shetrone2015} to find the best fit (through $\chi^{2}$ minimization) to the observed spectrum by varying T\raisebox{-.4ex}{\scriptsize eff}, surface gravity, microturbulence, metallicity, carbon abundance, nitrogen abundance, and $\alpha$-element abundance. In this analysis we use results from the 14th data release of SDSS (DR14, \citealt{dr142017,Holtzman2018,Jonsson2018}). 

To identify Sgr stream stars in the DR14 sample, we use the Sgr core sample from \citet{Hasselquist2017} that was selected using the methods described in \citet{Majewski2013} as a representation of the chemical abundance patterns of Sgr. This sample contains 158 stars with spectra having S/N $>$ 70 per half resolution element, which the APOGEE DR14 documentation defines as the lower threshold for which the APOGEE detailed chemical abundances are reliably measured. Because the bulk of the Sgr members are in the low-temperature regime (T\raisebox{-.4ex}{\scriptsize eff} $<$ 4200 K) where some elemental abundance determinations may exhibit systematic differences from higher temperature stars, we limit our MW sample in which we search for Sgr stream stars to the same stellar parameters of APOGEE Sgr stars. This sample is defined as:

\begin{itemize}
  \item 3550 $<$ $T_{eff}$ $<$ 4200
\item S/N $>$ 70 per half resolution element
\item No ``ASPCAPBAD'' flag set\footnote{This flag, described in detail in \citet{Holtzman2015}, indicates whether a star falls near edges of the synthetic spectra grids, or has a low $S/N$ spectrum.
}

\end{itemize}

Because the APOGEE Sgr core sample does not contain stars with [Fe/H] $<$ $-$1.2,  we also only consider MW stars with [Fe/H] $>$ $-$1.2. Distances to the APOGEE stars are derived by methods described in \citet{Hayden2014}, and are accurate to $\sim$ 20\%. These distances can be found as part of a value added catalog (VAC) from DR14\footnote{http://www.sdss.org/dr14/data\_access/value-added-catalogs/}. While the \emph{Gaia} distances from \citet{Bailer-Jones2018} are more precise at close distances, at d $>$ 5 kpc, the distance uncertainties become larger than $\sim$ 20\%, and the distances become prior-dominated. More than half of our APOGEE sample on which we perform the clustering analysis consists of stars with distances $>$ 5 kpc, so we adopt the spectro-photometric distances from the DR14 VAC to ensure our distance source is homogenous across the entire distance range we are able to probe with the APOGEE stars. 

\subsection{Gaia}

We supplement the APOGEE radial velocities and distance estimates with proper motions from \emph{Gaia} Data Release 2 \citep[DR2;][]{GaiaDR2}. Using the nearest neighbors matching tool within TOPCAT \citep{Topcat}, we identify the best \emph{Gaia} match (nearest neighbor within $1''$) for each APOGEE star. Of the 68 chemical candidates we identify in \S \ref{clust}, 62 have \emph{Gaia} proper motion matches. All but two of the sources matched within 0.4'' (the other two are at 0.73'' and 0.86''). We then use the full 6-D phase-space data (i.e., 3-D positions and velocities) to integrate orbits in a model Galactic potential, and further check the membership of our APOGEE-selected sample in the Sgr stream (described in further detail in \S \ref{res}).

To further investigate whether we have recovered true Sgr stream stars, we use the proper motions from \emph{Gaia} and the distances and radial velocities from APOGEE to calculate orbits of the Sgr stream candidates. Orbits were integrated in the \texttt{MWPotential2014} potential that is implemented in the \texttt{galpy} \citep{Bovy2015galpy} suite of Python routines\footnote{Available at \url{https://github.com/jobovy/galpy}}. This potential is a composite of a spheroidal bulge, a Miyamoto-Nagai disk, and an NFW halo (see default parameters in \citealt{Bovy2015galpy}). We note that while many recent works suggest that the halo mass of \texttt{MWPotential2014} may be too low to match Milky Way observations \citep[e.g., ][]{Helmi18,Fritz18,Watkins18}, we are mainly interested in the relative orbital properties of stars, so the absolute scale of the potential is unimportant. Throughout this work, we assume a circular velocity of 220~km~s$^{-1}$ at a Solar radius of 8.0~kpc from the Galactic center, and the Solar motion as measured by \citet{Schoenrich2010}.  We follow the \texttt{galpy} usage of a left-handed Galactic Cartesian coordinate (velocity) system with $X (U)$ positive toward the Galactic center, $Y (V)$ along Galactic rotation, and $Z (W)$ in the direction of the north Galactic pole. We initialize the orbit of each star at its current position and velocity, then integrate its orbit both forward and backward for 1.5~Gyr. From this, we estimate the orbital parameters, including the eccentricity, the peri- and apo-center distances ($r_{\rm peri}$ and $r_{\rm apo}$; relative to the Galactic center), and orbital energy and angular momentum.


\section{Selecting Sgr Stream Candidates in the APOGEE Data}
\label{clust}

\subsection{Kinematic Control Sample}
\label{sec:kin_control}

The goal of this work is to find Sgr members through chemistry alone, i.e., without any knowledge of kinematics or positions. However, to verify that the clustering algorithm used in this work is properly returning Sgr stream members by recovering the most obvious Sgr stream members, we select a ``kinematic'' control sample of likely Sgr stream stars from the APOGEE data. Sgr stream stars in the direction out of the MW plane can be easily distinguished from MW stars based on velocities and color. 

\begin{figure*}[h]
\includegraphics[width=1.0\columnwidth, trim=0.5in 0.5in 0.0in 0.5in]{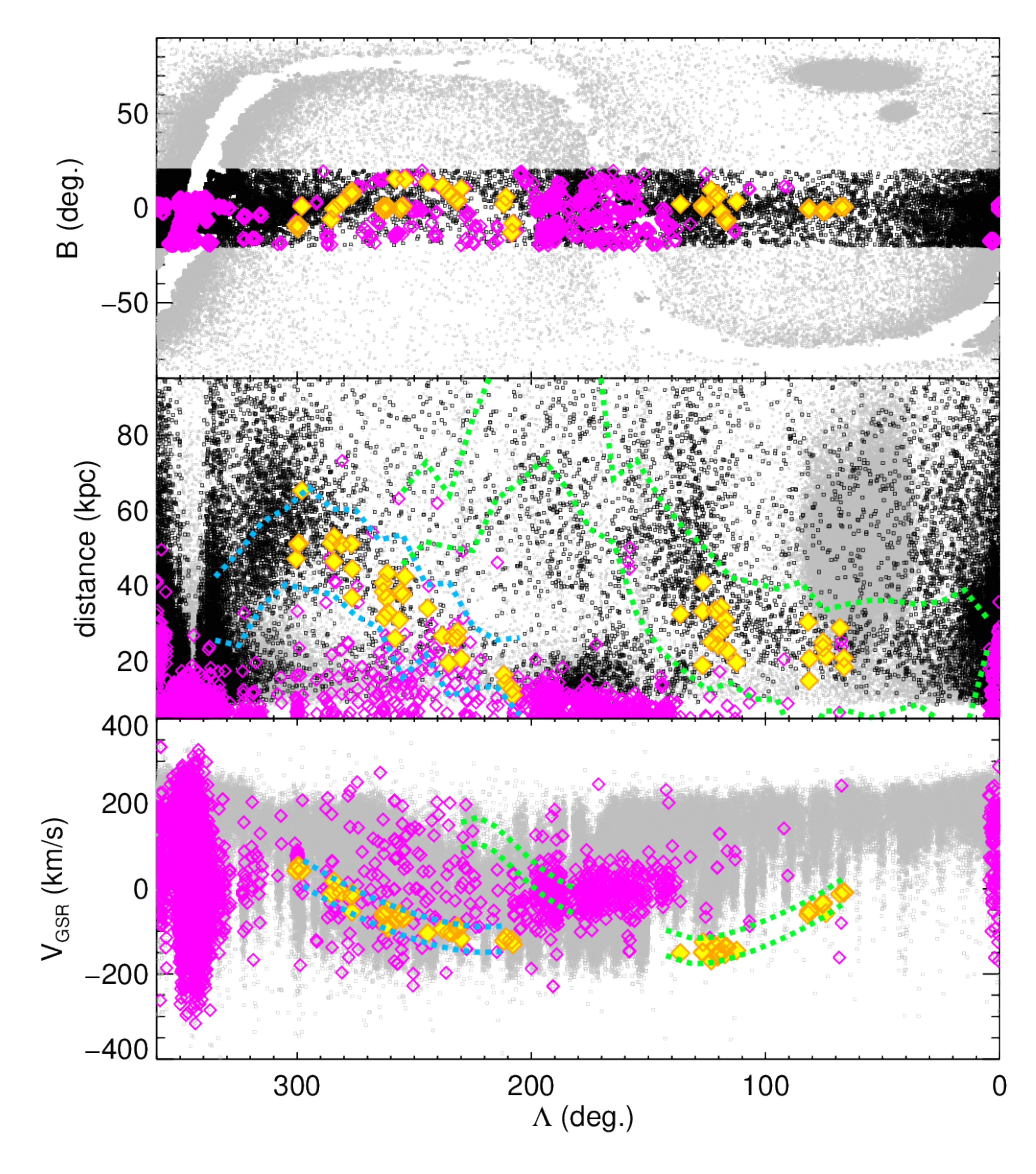}
\caption[Kinematic Sgr Sample Selection]{Illustration of our process to select Sgr stream candidates based on their kinematics. For comparison, the upper two panels show (as gray points) photometrically-selected M-giant candidates based on 2MASS+WISE color-color criteria (as in \citealt{Carlin2018}; see also \citealt{Koposov2014,Li2016a}); distances are estimated based on the photometric parallax relation derived for Sgr stream M-giants by \citet{Li2016a}. In the bottom panel, gray points are all APOGEE DR14 velocities. Small black points are a subset of the photometric sample selected based on position to be consistent with Sgr stream membership; the Sgr stream is clear in the distance panel (middle) as two swaths of black points arcing from $320^\circ \gtrsim \Lambda_\odot \gtrsim 200^\circ$ (leading arm) and $200^\circ \gtrsim \Lambda_\odot \gtrsim 30^\circ$ (trailing tail), mostly at distances beyond $\sim20$~kpc. Applying the criteria detailed in Section~\ref{sec:kin_control} to the APOGEE DR14 catalog yields the magenta sample in all three panels. Finally, we sub-select from the APOGEE sample using known velocity \citep{Belokurov2014} and distance \citep{Hernitschek2017} trends of the Sgr stream. Dashed lines represent the selections applied; each of these encompasses the $3\sigma$ range about the measured mean values. Our sample of kinematically selected Sgr candidates contains 58 stars, which are shown as yellow/orange diamonds in all three panels above.}\label{fig:select_cands}
\end{figure*}

The method for selecting the kinematic control sample is as follows:

\begin{enumerate}

\item We begin by applying cuts on distance, color, extinction, and proper motion to the entire APOGEE DR14 catalog (using distances from the value-added catalog's ``NMSU\_dist'' estimator). We transform the coordinates into the Sagittarius stream-aligned system devised by \citet{Majewski2003}.\footnote{In the coordinate system aligned with the Sagittarius stream, the coordinate $\Lambda$ is the angle along the stream, increasing along the trailing Sgr debris stream, and $\Beta$ is defined as the angular distance from the Sagittarius debris plane defined by 2MASS M-giants, analogous to Galactic latitude ($\Beta = 0^\circ$ is the Sagittarius debris plane).}, and select only stars with $\left|{\Beta}\right| < 20^\circ$ (must be within 20$^\circ$ of the Sgr debris plane). To avoid nearby disk stars, we require $d > 5$~kpc. We further limit the sample to red objects with $(J-K)_0 > 0.8$, with low total extinction $E(B-V) < 0.75$, and minimal proper motion (to remove nearby high proper motion stars) $\left|\mu_{\alpha \cos{\delta}}, \mu_{\delta}\right| < 15$~mas/yr. The resulting sample is plotted as magenta points in all three panels of Figure~\ref{fig:select_cands}. 

\item From this set of APOGEE candidates, we use the known properties of the Sgr stream to select candidates (also retaining only stars whose APOGEE spectra have $S/N>30$). The selections are based on velocity trends derived from SDSS members \citep[][dashed lines in the lower panel of Fig.~\ref{fig:select_cands}]{Belokurov2014}, and RR~Lyrae distances from PanSTARRS PS1 \citep[][dashed lines in center panel]{Hernitschek2017}. For each of these sub-selections, we include a range of $\pm3\sigma$ about the mean values from each study, using their reported $1\sigma$ errors. This yields a final sample of 58 kinematically-selected Sgr stream candidates, which are shown as yellow/orange diamonds in Figure~\ref{fig:select_cands}. 21 of these 58 stars have sufficient S/N and stellar parameters set by our criteria above, which we adopt as the ``kinematic'' sample. 

\end{enumerate}

This is not a complete sample of Sgr candidates from the APOGEE DR14 database, since the selection criteria are limited to regions with previous measurements. However, it is a reliable set of candidates to use as a test of our chemical tagging selection. We have confirmed that these selection criteria recover the 42 Sgr stream stars observed at high spectral resolution in the optical by \citet{Carlin2018}. 

\subsection{K-means Clustering}

\begin{figure*}[h]
\epsscale{1.0}
\plotone{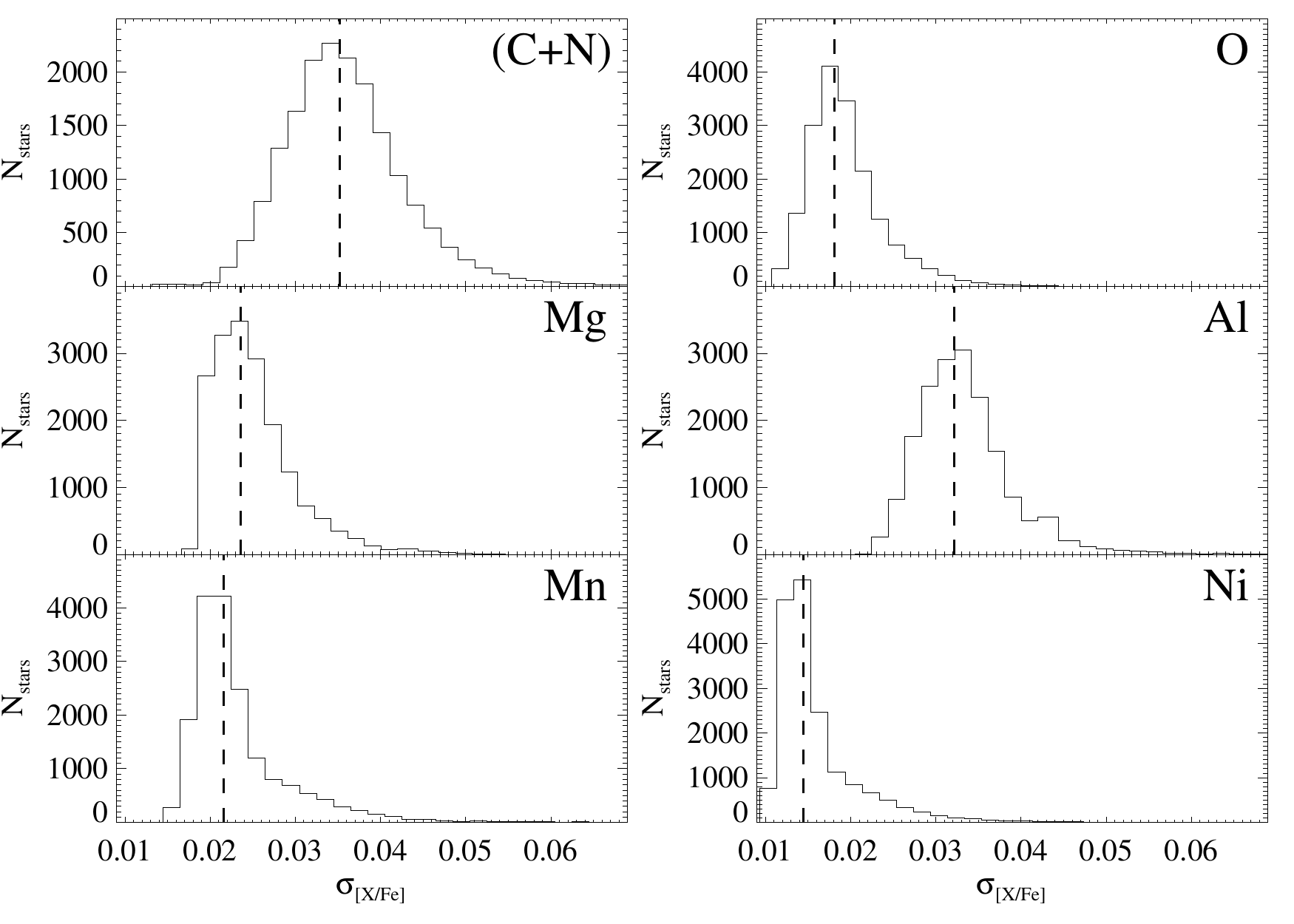} 
\caption[uncertainty]{Distributions of the APOGEE chemical abundance uncertainties for the sample on which we perform our clustering analysis. Dashed lines mark the median uncertainty, which we adopt as weights in the \emph{k} means clustering analysis.}
\label{fig:uncert}
\end{figure*}

To conduct our search for Sgr stream members in chemical abundance space, we use the \emph{k}-means ``CLUST\_WTS'' function\footnote{http://www.harrisgeospatial.com/docs/CLUST\_WTS.html} in IDL. This function randomly chooses the starting points of \emph{k} clusters and then moves stars in and out of clusters to minimize variance within each cluster and maximize variance between clusters. We set this function to run with 100 different random starting guesses, which we find consistently returns the same clusters for \emph{k} $<$ 50. We choose the APOGEE elemental abundances in which Sgr exhibits clear deficiencies, and have low reported uncertainties from the APOGEE pipeline ($\sim$ 0.05 dex or less). The $N$-dimensional space in which we cluster is then defined by [O/Fe], [(C+N)/Fe], [Mg/Fe], [Al/Fe], [Mn/Fe], and [Ni/Fe]. The uncertainties for each element are shown in Figure \ref{fig:uncert}. The median uncertainty of each element are used as weights in the \emph{k}-means clustering analysis such that elements with lower uncertainties, such as Ni and O, are given larger weight than elements with higher uncertainties, such as Al and C+N.

\subsection{Selecting \emph{k} clusters}

With a kinematic sample of expected stream members and the core sample defining the abundance patterns of Sgr, we determine the number of clusters, \emph{k}, in which to separate the data. To select an appropriate value for \emph{k}, we run the clustering algorithm for multiple values of \emph{k} from 3 to 50. For each value of \emph{k}, we track the Sgr core sample from \citet{Hasselquist2017} and the kinematic control sample defined above. The cluster that contains the bulk of the Sgr core sample is referred to as the ``Sgr cluster'', and should contain the majority of the kinematic control sample if we are to identify potential stream members. We choose the ideal value of \emph{k} to be the number of clusters that results in a marginal splitting (10\%) of the Sgr core sample from the Sgr cluster. In this way, we are ``tuning'' the \emph{k}-means algorithm.

\begin{figure*}[h]
\epsscale{1.0}
\plotone{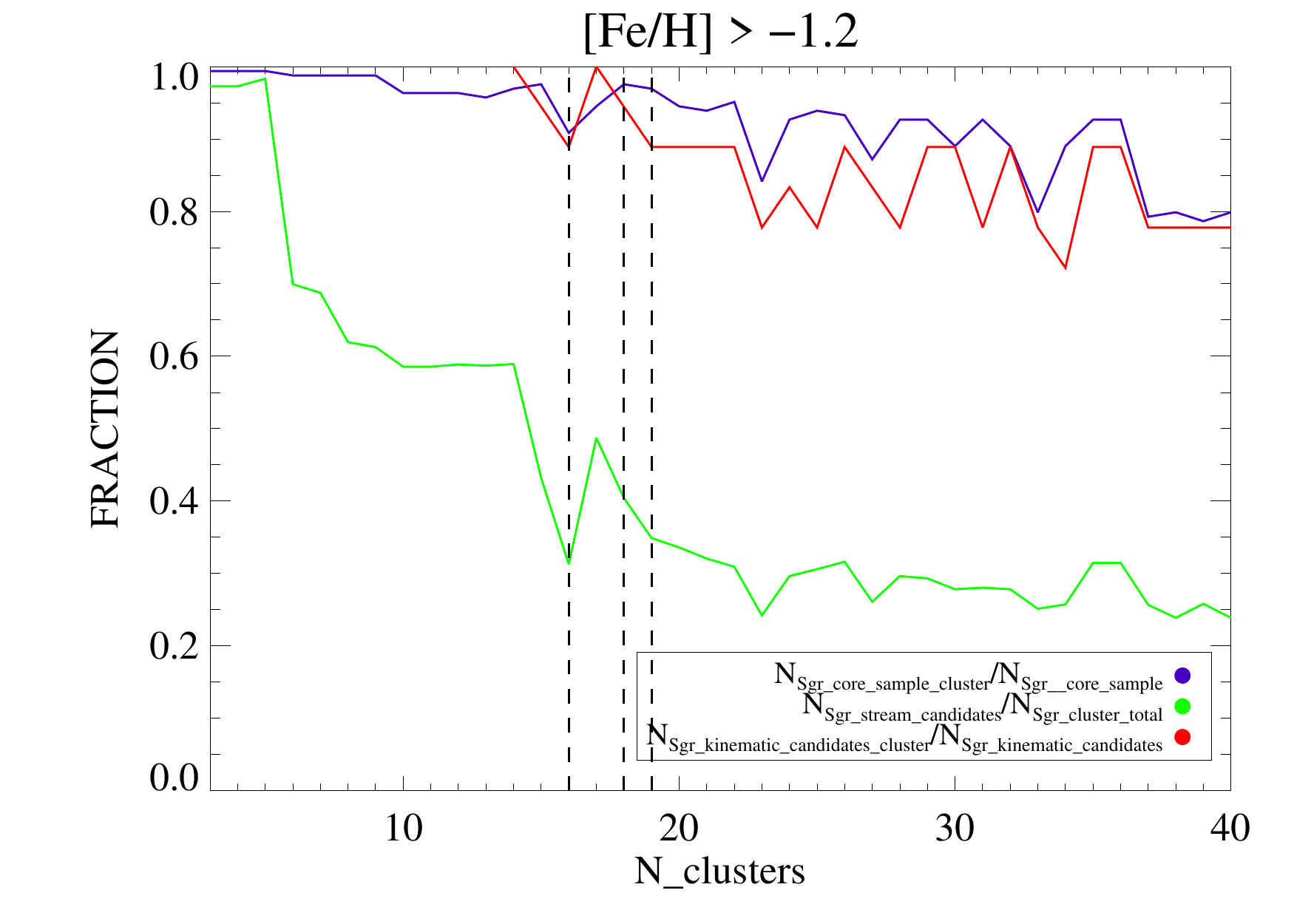} 
\caption[Cluster Behavior as a Function of \emph{k}]{Ratios of stellar samples plotted as a function of \emph{k} clusters. The lines are colored according to the legend in the lower-right. The dashed lines mark best-choice values for \emph{k}, defined as where 10\% of Sgr core sample separates from the ``Sgr cluster''. In this work, we present results for \emph{k} = 16.}
\label{fig:cluster_stats}
\end{figure*}

The motivation for our choice of \emph{k} is shown graphically in Figure \ref{fig:cluster_stats} where we plot the fraction of stars retained in the Sgr cluster as a function of \emph{k} for each sample. The blue line, which represents the fraction of Sgr core members retained in the Sgr cluster, exhibits a decrease at \emph{k} = 16 which corresponds with a similar decrease in the fraction of kinematic sample retained (red line). We adopt \emph{k}=16 as our best number of clusters in which to separate the data, but note that 18 and 19 are just as viable choices based on our criteria set here. However, these choices result in very little difference in the amount of  ``chemical'' stream candidates; the fraction of the Sgr cluster that is made up of stream candidates (stars that aren't Sgr core members) is denoted by the green line in Figure \ref{fig:cluster_stats}.

\section{Results}
\label{res}

\begin{figure*}[h]
\epsscale{1.0}
\plotone{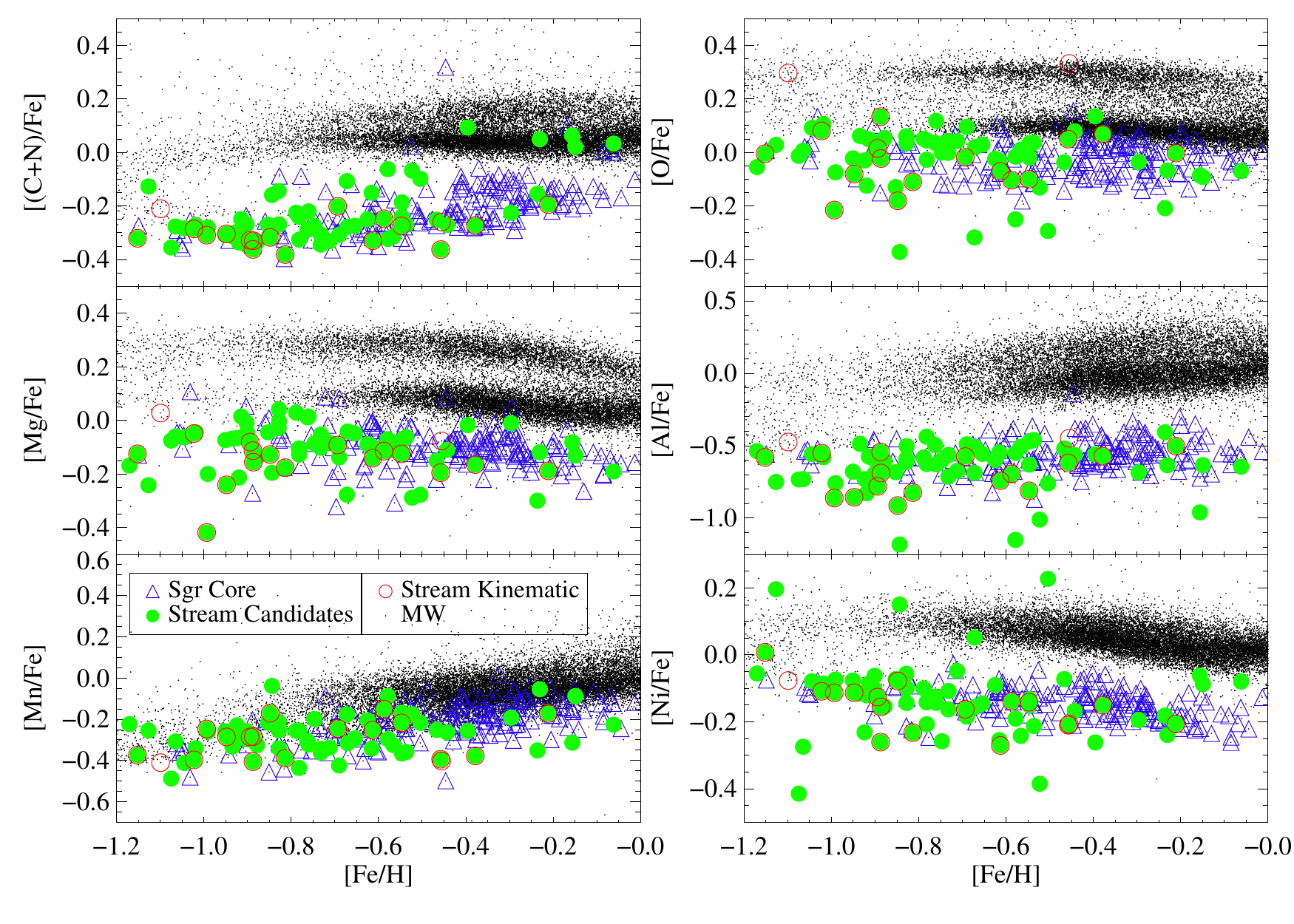} 
\caption[Sgr Stream Chemical Abundance Projections]{Two-dimensional chemical abundance projections for the APOGEE sample on which the clustering algorithm was performed. Points are colored according to the legend in the lower-left panel.}
\label{fig:chem}
\end{figure*}

Performing our clustering analysis with \emph{k} = 16 results in a cluster containing 90\% of the Sgr core sample along with an additional 68 stars that do not reside in the Sgr core. 19 of the 21 kinematic candidates selected in \S \ref{sec:kin_control} are in this sample of 68, demonstrating that our method reliably selects Sgr stream stars. These 68 stars are henceforth considered to be ``Sgr stream candidates''. The chemical abundance patterns for the Sgr core and Sgr stream candidates are shown in Figure \ref{fig:chem}, along with the chemical abundance patterns of the entire MW sample on which the clustering algorithm was performed.

The Sgr stream candidates fall in the same regions of these 2D abundance projections as the Sgr core sample, with the exception of a handful of cases as we now describe. There are 4 stars with [Fe/H] $>$ -0.25 that exhibit [(C+N)/Fe] abundances consistent with the MW trend, unlike the majority of Sgr core stars at that metallicity. However, there are 3 Sgr core stars that also occupy this region of [(C+N)/Fe]-[Fe/H] abundance space. There are also 4 stars with enhanced [Ni/Fe] that do not follow the bulk Sgr core trends. 

The Sgr stream candidates are generally more metal-poor than the Sgr core, which is consistent with previous abundance studies of the Sgr stream (see, e.g., \citealt{Chou2007,Carlin2018}). In fact, the stream has stars more metal-poor than our cutoff of [Fe/H] $>$ $-$1.2 (see, e.g., \citealt{Chou2007}), so we are likely missing metal-poor Sgr stream members that reside in the APOGEE data (discussed further in \S \ref{sec:disc}). 

\begin{figure*}[!t]
\includegraphics[width=0.49\textwidth]{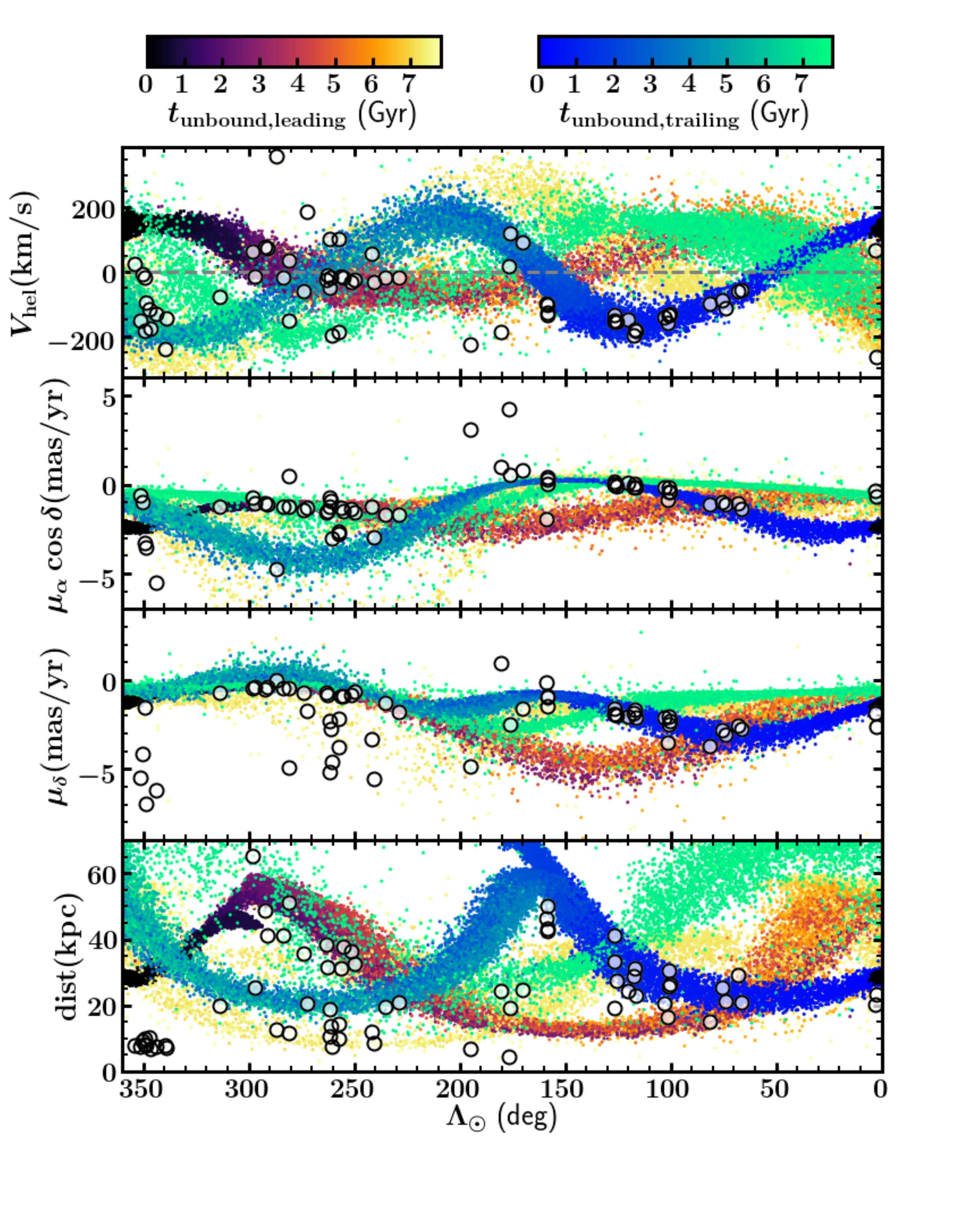}
\includegraphics[width=0.49\textwidth]{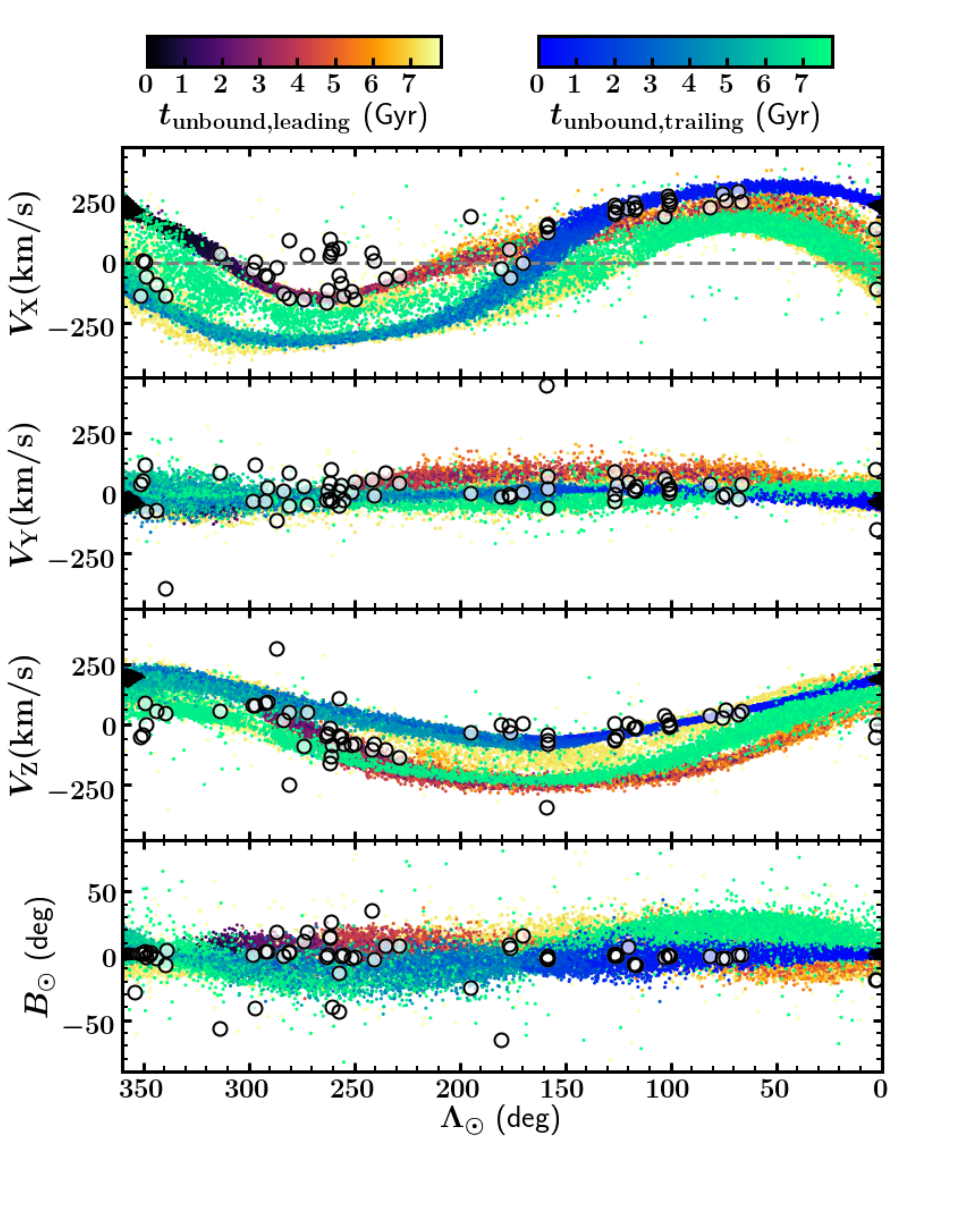}
\caption[Proper motion comparison to LM10]{Positions and velocities of the Sgr stream candidates (white circles) compared to the predicted positions and velocities of the \citet{Law&Majewski2010} model of the Sgr system (colored dots). Left: Comparisons of the APOGEE radial velocities, \emph{Gaia} proper motions, and APOGEE distances plotted as a function of $\Lambda_\odot$ . Right: Comparisons of calculated UVW velocities and $\left|{\Beta}\right|$ as a function of $\Lambda_\odot$. The \citet{Law&Majewski2010} points are colored according to whether they belong to the leading/trailing debris, as well as when they were stripped from Sgr, as indicated by the color bar above.}
\label{fig:model_comp}
\end{figure*}

Using the Gaia proper motions, we compare the positions and velocities of the Sgr stream candidates to those predicted for Sgr tidal debris from the \citet{Law&Majewski2010} model of the Sgr system. This N-body model fits the positions and kinematics of the most recent Sgr tidal debris by introducing a non-axisymmetric component to the MW gravitational potential. The comparisons are shown in Figure \ref{fig:model_comp}.  Sgr stream candidates are plotted as white circles superposed on the model debris from \citet{Law&Majewski2010}, which are colored by when the debris was stripped from Sgr. From these comparisons, we find that with the exception of a group of stars clearly residing in the bulge ($\Lambda_\odot$ $\sim$ $350^\circ$ and d = 8 kpc), most of the stream candidates have positions and velocities consistent with those predicted by the model. One obvious exception is the group of stars with 240 $< \Lambda_\odot <$ 290 with U (i.e., $V_{\rm X}$) velocities that are higher than those predicted by the model.

\begin{figure*}[h]
\includegraphics[width=0.49\textwidth]{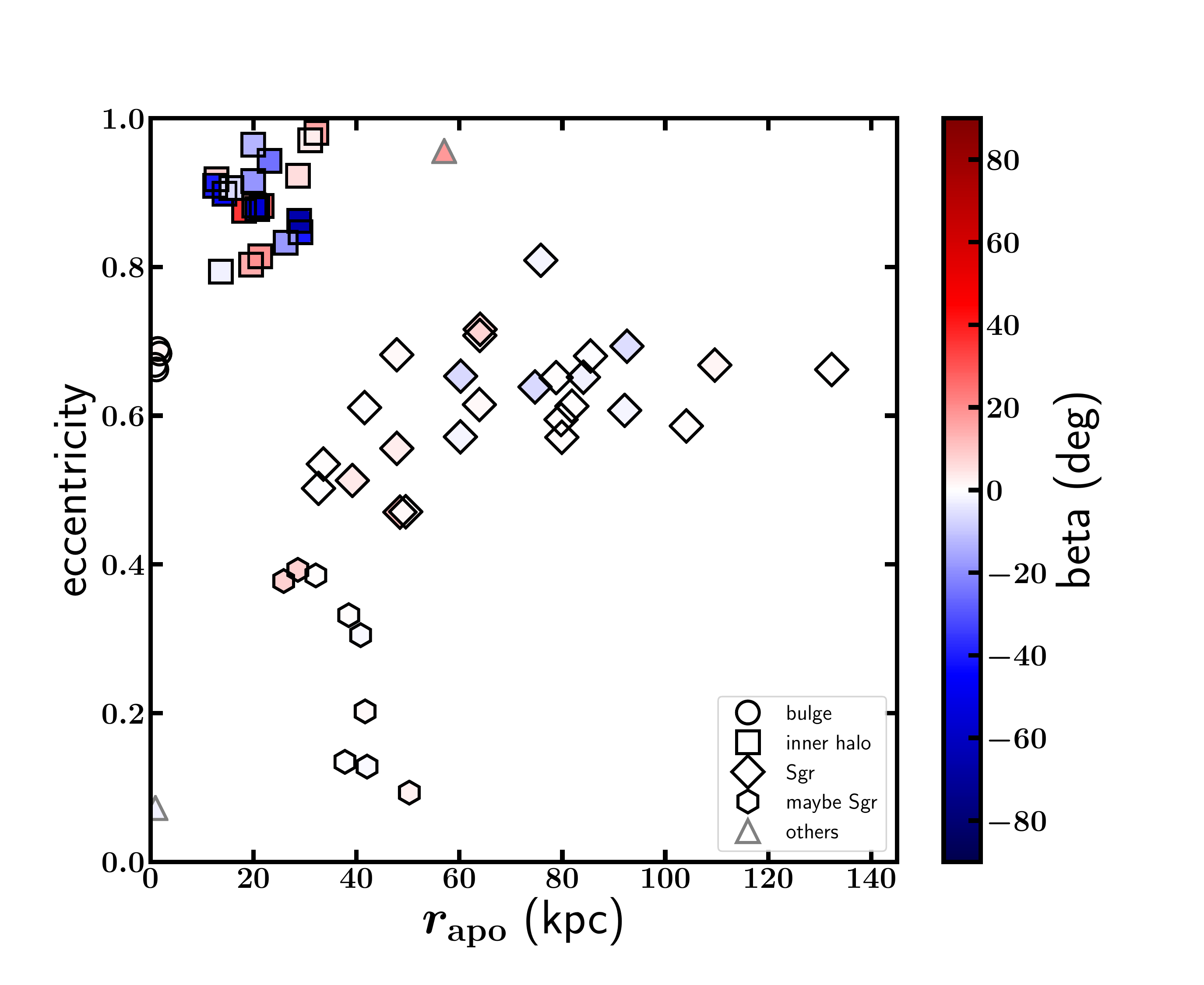}
\includegraphics[width=0.49\textwidth]{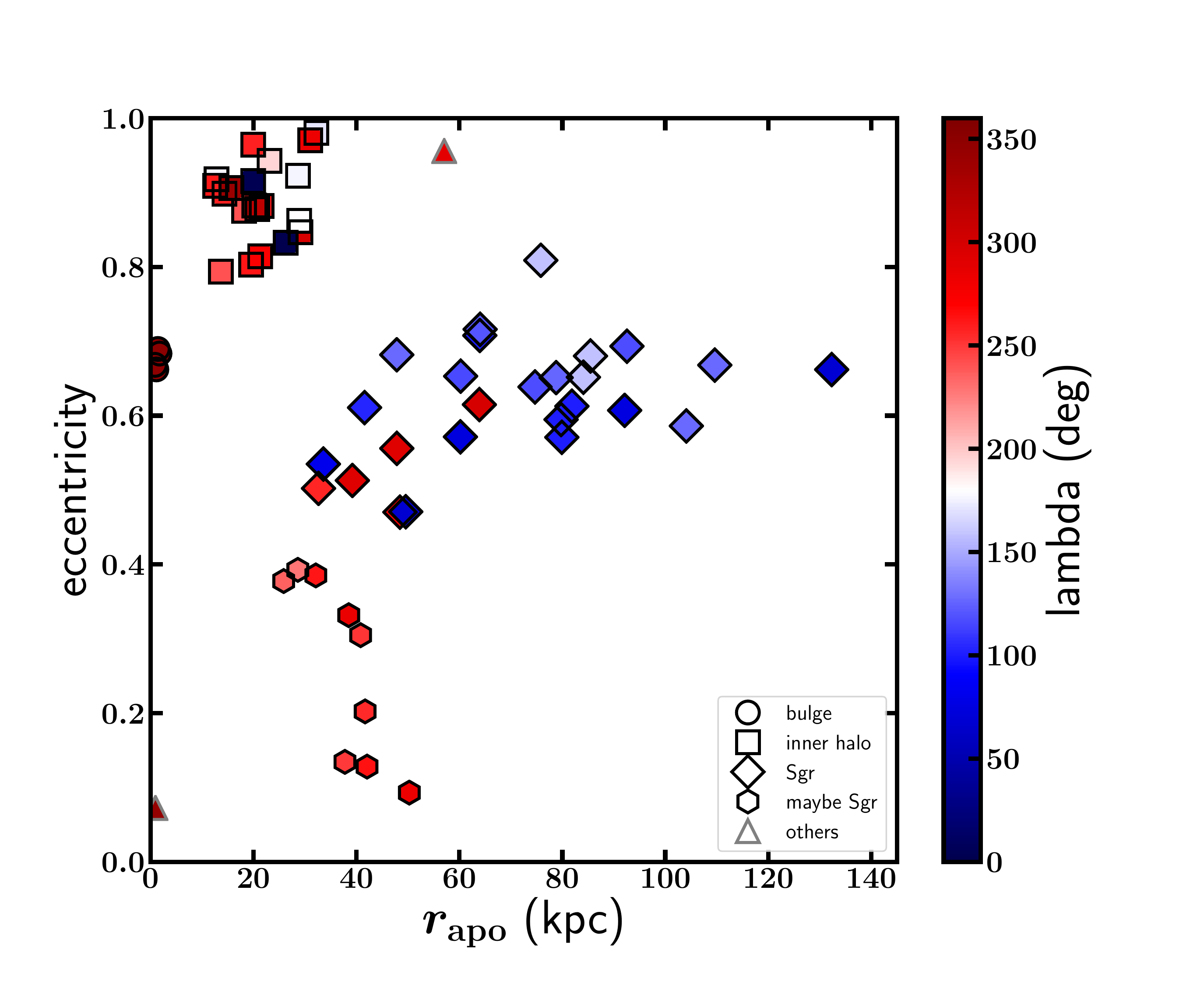} 
\includegraphics[width=0.49\textwidth]{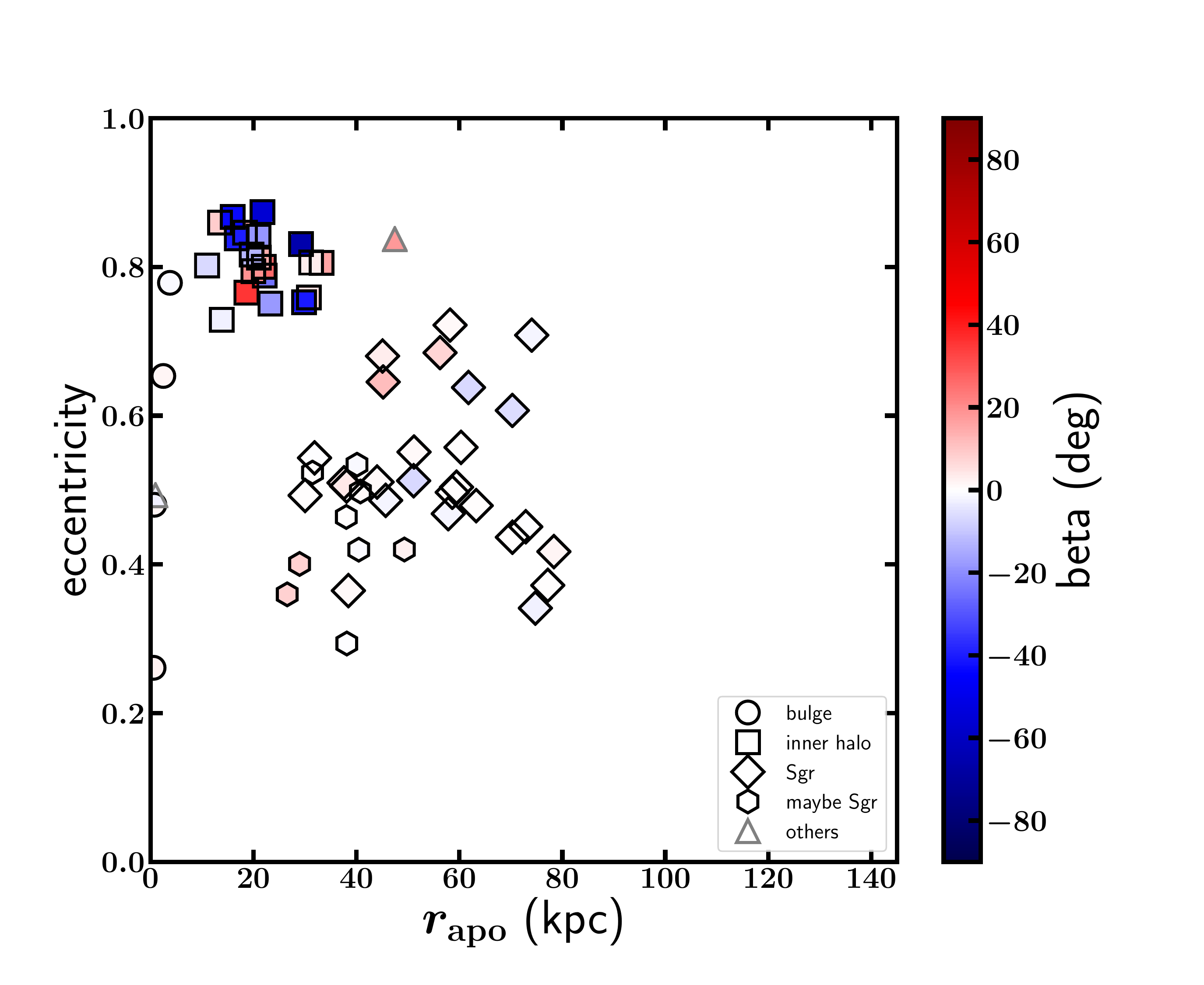}
\includegraphics[width=0.49\textwidth]{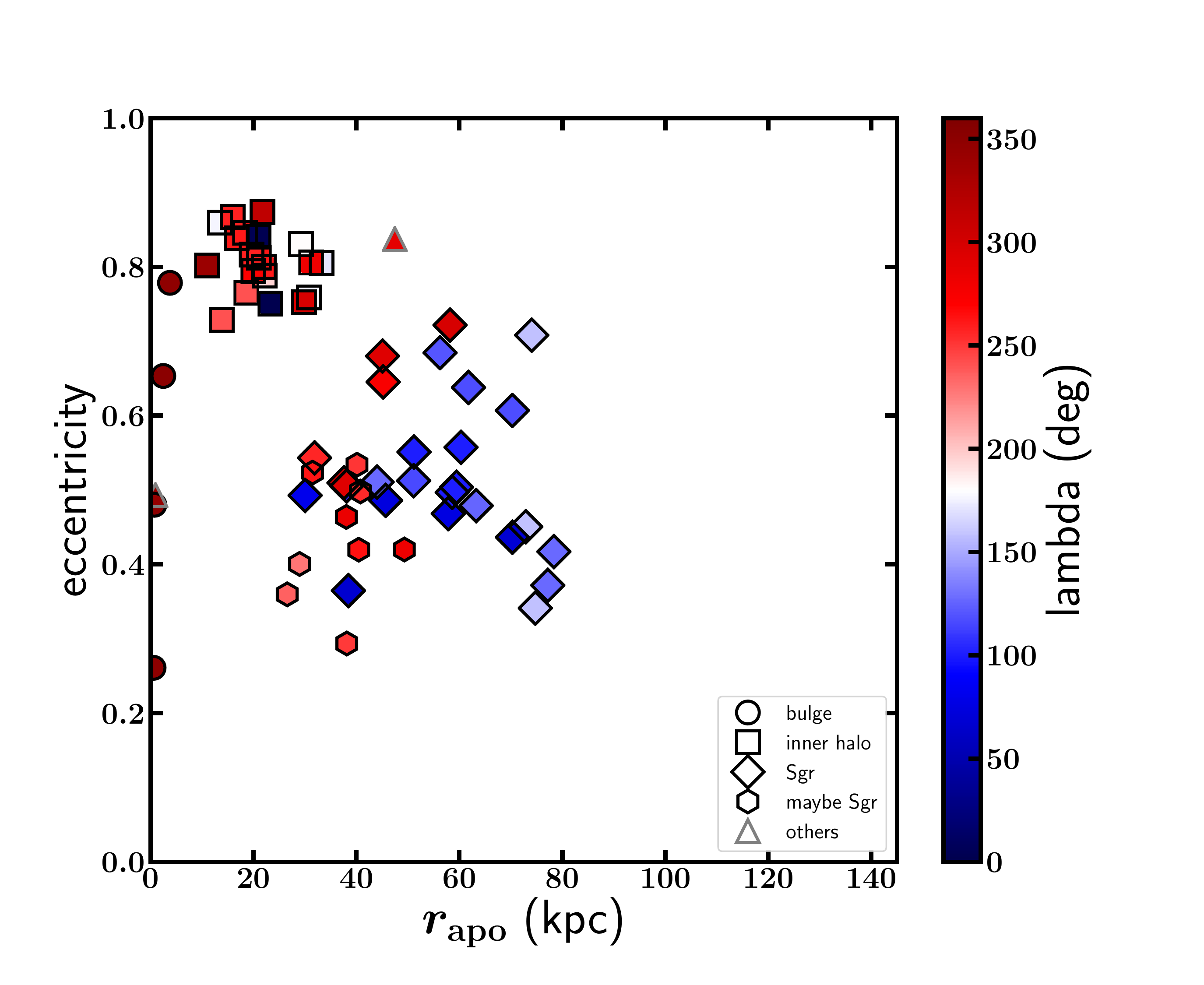} 
\caption[Comparison to LM10 Model]{Calculated Orbital eccentricity plotted as a function of orbit apocenter, colored by $\left|{\Beta}\right|$ (left) and $\Lambda_\odot$ (right). The point symbols correspond to which orbital group the stream candidates belong to, as described in the text, and indicated in the legend. The top row shows orbital results using the MWPotential2014 gravitational potential, and the bottom row shows orbital results using the gravitational potential used in \citet{Law&Majewski2010}.  }
\label{fig:orbit}
\end{figure*}

In top row of Figure \ref{fig:orbit} we plot the orbital eccentricity as a function of orbit apocenter ($r_{\rm apo}$) as calculated in the \texttt{MWPotential2014} gravitational potential for the chemically selected Sgr stream candidates. In the left column, the points are colored by the Sgr $\Beta$ coordinate, such that the color saturation corresponds to the distance from the Sgr debris plane; stars with white or light-colored shading are near the Sgr orbital plane, while those that are darker red or blue are far from the Sgr plane. In the right column the points are colored by the Sgr $\Lambda_\odot$ coordinate, such that the blue and red points indicate whether the stars have $\Lambda_\odot$ consistent with Sgr trailing arm or leading arm debris, respectively.

Analyzing the stars in this way, we find that the Sgr stream candidates split into 3 orbital groups along with one hypervelocity star (HVS). The catalog of APOGEE IDs and membership assignments is shown in Table \ref{tab:stream_cand}. 
\begin{itemize}
  \item Sgr Stream (35 stars) --- This includes 26 stars (diamonds in Fig.~\ref{fig:orbit}) with eccentricities between 0.4-0.85 and apocenters between 25-140~kpc that correspond to typical orbital properties of Sgr debris. The 9 additional points shown as hexagons in Fig.~\ref{fig:orbit} are on more circular orbits than the main Sgr sample, but are all near $\Beta = 0^\circ$ and have apocenters consistent with Sgr debris. We call these ``maybe Sgr,'' but believe they are likely to be related to the stream.
  \item Accreted Halo (20 stars) --- A distinct group of stars (shown as squares in Fig.~\ref{fig:orbit}) that have high eccentricities ($> 0.75$) and apocenters of $\sim$15-30 kpc stands out in Figure~\ref{fig:orbit}. Note that stars in this category appear on both sides of the Sgr plane, and are mostly located at $\Lambda > 200^\circ$ in the North Galactic cap. 
  \item Bulge (4 stars) --- Four stars (circles in Fig.~\ref{fig:orbit}) are on eccentric orbits, but with very small apocenters, as is typical of stars in the Galactic bulge.
  \item HVS (1 star) --- This star does not appear on this plot because it has no measured apocenter (at the last time step of our orbit integrations, it is $\sim760$~kpc from the Galactic center) -- its observed velocities suggest it is unbound in \texttt{MWPotential2014}.
  \item Others (2 stars) --- Two stars do not fit into any of the groups we identified above; we classify these simply as ``others'' and do not consider them further.
\end{itemize}

As previously noted, the choice of gravitational potential has little to no effect on our results. We classified stars based on their \textit{relative} orbital parameters; groupings in parameter space such as eccentricity vs. $r_{\rm apo}$ are likely to be similar even when orbits are integrated in different potentials. Nonetheless, because we compare to the \citet{Law&Majewski2010} model in Figure \ref{fig:model_comp}, we integrate orbits in the potential that was used in that work. To do so, we use the \texttt{gala}\footnote{Available at \url{http://gala.adrian.pw}.} software (\citealt{PriceWhelan2017a,PriceWhelan2017b}), which has explicitly implemented the \citet{Law&Majewski2010} potential. Orbits integrated in the \citet{Law&Majewski2010} gravitational potential yield the results in the lower panels of Figure \ref{fig:orbit}. The different shapes in these panels are based on the classifications from the \texttt{MWPotential2014} results in the upper panels. While the positions of individual stars shift between the upper and lower panels in Figure \ref{fig:orbit}, the classifications are retained. In fact, it is encouraging to see that the ``maybe Sgr'' sample overlaps the ``Sgr stream'' sample more when integrated in the LM10 potential. Finally, we note that while the uncertainties on most stars' orbits are rather large, the separation between populations in Figure \ref{fig:orbit} is large enough that errors would have little effect on our classifications.

\begin{table*}[ht]
  \caption{Sgr Stream Cluster}
\begin{center}
\begin{tabular}{l c c c c c c r}
\hline
  \textbf{APOGEE ID} & \textbf{Membership\tablenotemark{*}} & $r_{\rm apo}$ (kpc) & $\sigma^{+}_{r_{\rm apo}}$ & $\sigma^{-}_{r_{\rm apo}}$  & $e$ & $\sigma^{+}_{e}$ & $\sigma^{-}_{e}$\\
\hline
2M00151324-1430281  & a    & 60.20   &   28.70     & 15.87     &  0.57    &   0.08      & 0.05\\
2M00194682-1345014   &a     & 92.10   &  165.03    &  49.75  &     0.61  &     0.20    &   0.17\\
2M00445288-1244488  & a      &33.57    &   6.87     &  4.83   &    0.53    &   0.02    &   0.00\\
2M01512785-0352240   &a     & 79.75   &  112.74    &  32.76    &   0.59    &   0.15   &    0.03\\
2M01532039-0327406  & a      &81.86   &  120.18    &  34.69    &   0.61    &   0.15    &   0.03\\
2M01535096-0320382   &a     & 79.90    & 127.77    &  33.77    &   0.57    &   0.16    &   0.01\\
...\tablenotemark{**} \\
\tableline
\end{tabular}
\\
\raisebox{+.4ex}{\scriptsize *} a = Sgr stream, b = accreted halo, c = bulge, d = HVS, e = others \\
\raisebox{+.4ex}{\scriptsize **}The full list is available from the online journal. \\
\end{center}
\label{tab:stream_cand}
\end{table*}

\begin{figure*}[h]
\epsscale{1.0}
\plotone{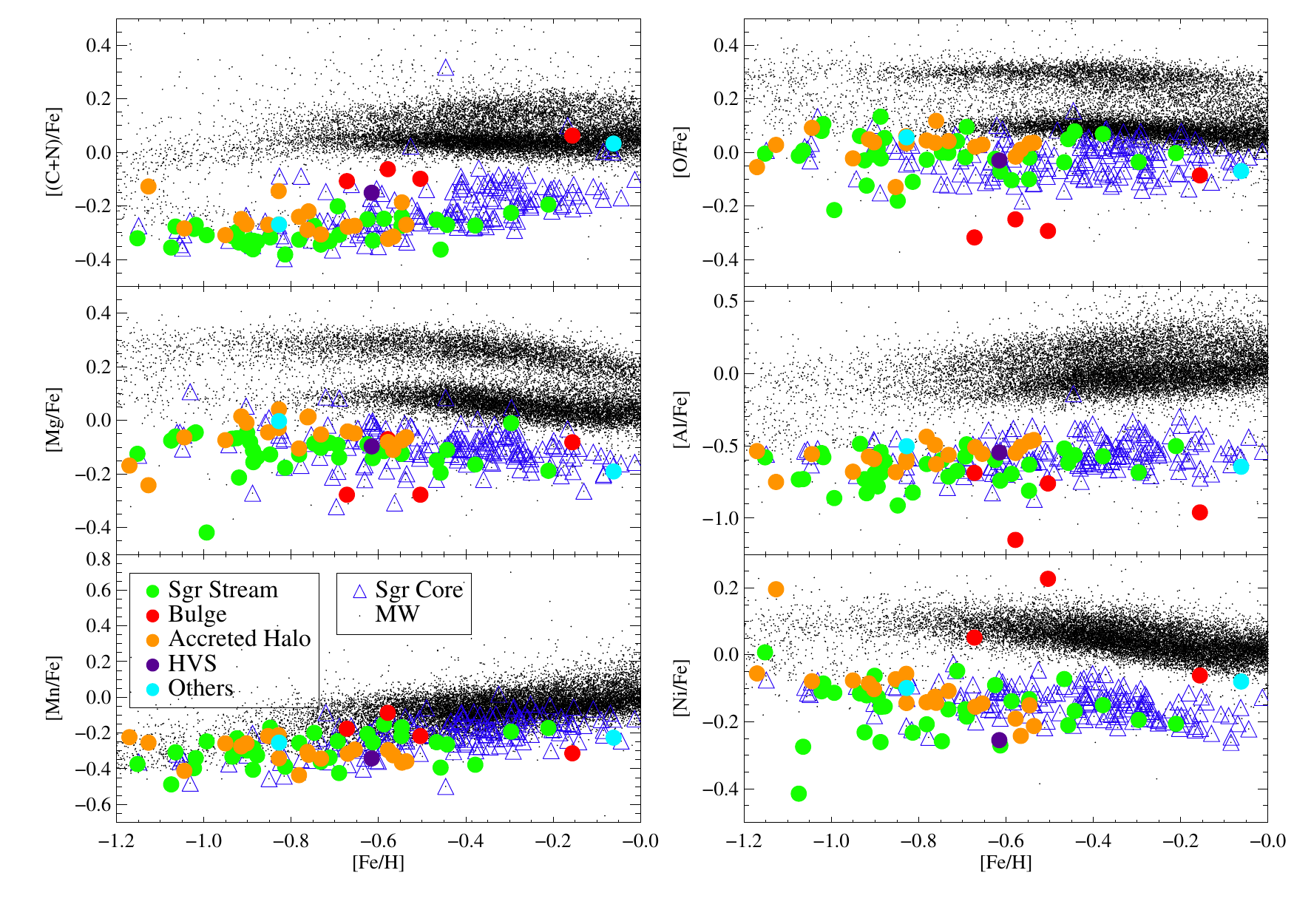} 
\caption[Chemical Abundance Projections for Sgr Stream Sub-groups]{Same as in Figure \ref{fig:chem} except the Sgr stream candidates are now colored according to the orbital group defined in the text (see the legend in the lower-left panel).}
\label{fig:chem_label}
\end{figure*}

In Figure \ref{fig:chem_label} we plot again the 2D abundance patterns, this time coloring the points based on the orbit divisions described above. We find that the apparent chemical outliers mentioned above appear to reside in the bulge sample. In the following sections, we describe each sample in more detail and comment on whether or not they can be associated with the Sgr system. 

\subsection{Sgr Stream}

We find that 35 stars (including 19 stars from the kinematic control sample) of the 62 chemical candidates with \emph{Gaia} proper motions follow the expected positions and kinematics of the \citet{Law&Majewski2010} model of the system. Of these, 21 stars belong to the trailing arm and 14 stars belong to the leading arm, including all $\sim9$ of the low-eccentricity ``maybe Sgr'' stars. This is indicated by the coloring in the top-right panel of Figure \ref{fig:orbit}.
According to this model, these stars were all stripped in the most recent pericenter passage. In Figure \ref{fig:orbit} we find that the trailing debris exhibit a range of apocenters from 60-120 kpc, and the leading arm debris have apocenters of 40-60 kpc. These are more in line with the apocenters found by \citet{Belokurov2014} and \citet{Fardal2018} ($\sim$ 48 kpc and $\sim$ 100 kpc, respectively). 

We find Sgr stream stars as metal-rich as [Fe/H] = $-$0.2, with the majority of stream stars exhibiting [Fe/H] $<$ $-$0.6. The mean metallicity of our Sgr stream sample is more metal-poor than the Sgr core sample, suggesting that there was a radial metallicity gradient across the Sgr progenitor when it began merging with the MW such that the more metal-poor stars were stripped first. 





\subsection{Accreted Halo}

In Figure \ref{fig:orbit}, we define the group of stars that have highly eccentric orbits and median $r_{\rm apo} = 20.2 $~kpc as the ``accreted halo'' stars. These stars are similar to the accreted halo stars described in the literature (e.g.,  \citealt{Schuster2012,Belokurov2018,Deason2018}), which concludes that the pileup of metal-rich ([Fe/H] $>$ -1.5) halo stars at apocenters of $\sim20-25$~kpc originates from a major merger some 8-11 Gyr ago. They argue that a massive satellite interacting with the disk can deposit stars on highly radial orbits (see, e.g., \citealt{Belokurov2018}). Further support for this interpretation can be found in the literature (see, e.g., \citealt{Helmi2018,Kruijssen2018,Mackereth2019}, and \citealt{Chiba2000,Brook2003} for earlier reports), although we explore other potential origins of these stars in \S \ref{sec:halo}. 

The accreted halo stars are unlikely to be Sgr stream stars as they generally have $\left|{\Beta}\right|$ values that put them well outside the predicted plane of Sgr debris. However, by the nature of our analysis, these stars have the same chemistry as the Sgr stream stars. This suggests that the dwarf galaxy that merged with the MW some 8-11 Gyr ago may have undergone a very similar star formation history as Sgr, at least until an enrichment of [Fe/H] $\sim$ $-$0.6, the highest metallicity among ``accreted halo'' stars.

\subsection{Bulge}

There are four Sgr chemical candidates that actually reside in the Galactic bulge and are on bulge-like orbits. Thus, they are unlikely to be actual stream members. We find that by increasing \emph{k}-clusters, we cannot remove these stars from the Sgr chemical candidate cluster without removing likely stream members, suggesting a potential limitation to using the \emph{k}-means technique to conduct chemical tagging analysis. 

\begin{figure*}[h]
\epsscale{1.0}
\plotone{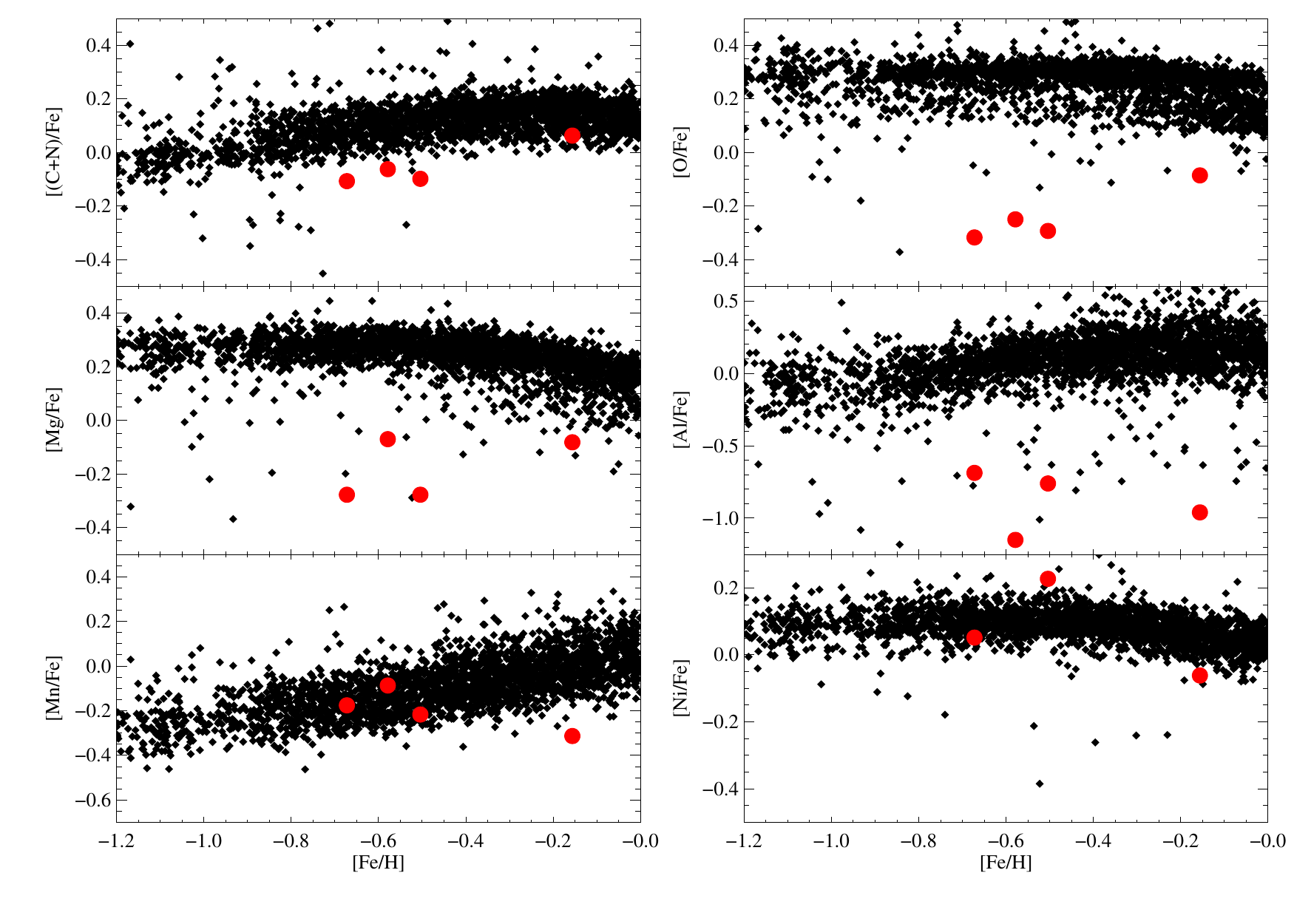} 
\caption[Bulge Chemical Comparison]{Two-dimensional chemical abundance plots of the bulge stars that were tagged as Sgr chemical candidates (red) compared to a bulge subsample of our MW comparison sample selected as MW stars having R $<$ 3 kpc (black).}
\label{fig:bulge_chem}
\end{figure*}

These stars are likely grouped with Sgr stream stars due to their exceptionally deficient [O/Fe] abundance ratios, as well as deficient [Mg/Fe] and [Al/Fe], as seen in Figure \ref{fig:chem_label}. They may not be Sgr stream stars, but they also are not typical bulge stars. In Figure \ref{fig:bulge_chem} we compare the bulge stars (red) to a strictly ``bulge'' MW sample, which is a subsample of our previous MW comparison sample (MW stars with the same parameters as the Sgr core sample), but with $R_{Gal}$ $<$ 3 kpc. While there are some bulge stars with similarly deficient abundance ratios, the bulge Sgr stream candidates are deficient by $\sim$ 0.5 dex in [O/Fe], by $\sim$ 0.4 dex in [Mg/Fe], and by $\sim$ 0.7 dex in [Al/Fe] compared to the bulk of the bulge stars. They appear to have the same [Mn/Fe], and slightly deficient in [(C+N)/Fe].

The bulge Sgr stream chemical candidates actually have [O/Fe] $\sim$ 0.2 dex lower than the actual Sgr stream candidates. They are also enhanced in Ni and Mn relative to the Sgr stream candidates, with Ni and Mn abundances that more closely track the other stars in the bulge. Two of these stars actually have enhanced [Ni/Fe] relative to the bulge stars. This suggests that these may be some class of stars that formed from material containing excess Type Ia SNe yields as compared to the majority of the MW bulge stars. 

While determining the origin of these stars is beyond the scope of this work, the presence of these stars in the Galactic bulge suggests that either there was an epoch of star formation in the bulge that formed stars from gas that was unusually enriched in Type Ia SNe, or the MW underwent a merger of a dwarf galaxy that was as metal-rich as Sgr, but formed stars from gas that had a higher ratio of Type Ia/Type II ejecta than Sgr.

\subsection{HVS}

One star in the Sgr chemical sample is at a Galactocentric distance of $\sim54$~kpc, near the position of the most recent apocenter of the LM10 Sgr orbit, which happened $\sim450$~Myr ago. Contrary to expectations for tidal debris at apocenter, this star has a total Galactocentric space velocity of $\sim577^{+209}_{-129}$~km~s$^{-1}$. In fact, an orbit integration suggests that this star's closest approach will occur in $\sim20$~Myr, at $r_{\rm peri} \sim 53$~kpc, and that it is completely unbound from the Milky Way. If this hypervelocity star's orbit does not trace back to the Galactic center, then what might its origin be? The backwards integration of its orbit carries it from its current location near the Galactic anticenter in the 3rd quadrant of the North Galactic Hemisphere, meaning that its orbit passes nowhere near the LMC or M31. Thus, if this star is an interloper from beyond the MW, its origin will require further study. 

Given that this HVS candidate is an early M-giant, it is unlikely to originate as a ``runaway'' in a binary ejection scenario \citep[e.g., ][]{Blaauw1961}, since this formation mechanism requires young, massive stars. We have already ruled out scenarios involving ejection from a black hole at either the center of the Milky Way, M31, or the LMC. Thus, the most likely origin of this distant HVS is a dynamical interaction (e.g., between the components of a triple system, or between a binary and a passing massive star; e.g., \citealt{LeonardDuncan1990,Gvaramadze2009}). However, these mechanisms often require a very massive star to perturb the system; given the old, metal-poor nature of the Sgr debris (and the surrounding halo stars) in the region of sky where the HVS candidate is located, this mechanism also seems unlikely to have created this HVS candidate. 

There is some support in the literature for HVSs originating from mergers. \citet{Abadi2009} find that disrupting dwarf galaxies can contribute HVSs; however, the mechanism requires that the HVSs were stripped from their host dwarf on a recent pericentric passage (i.e., near the Galactic center). Given that the star we have found never goes within $\sim50$ kpc of the Galactic center, this scenario is unlikely. Similarly, \citet{Piffl2011} suggest that massive dwarf galaxy mergers (M $> 10^{9} M_{\odot}$) can result in a population of HVS that is essentially the high-velocity tail of the debris. However, at the apocenter of the Sgr orbit, where we find the HVS candidate, the mean Galactocentric velocity should be roughly zero. A high-velocity tail relative to the mean velocity is unlikely to produce a star with velocity $>600$~km~s$^{-1}$. Again, the simulations discussed by \citet{Piffl2011} require the stars to be debris stripped on the most recent pericentric passage, which seems to be ruled out because our candidate HVS\'s orbit does not trace back along the Sgr orbit.

%

It is possible that this candidate HVS simply has a systematic error in its proper motion, which contributes most of the star's 3D space velocity (though its $v_{\rm gsr} = -111$~km~s$^{-1}$ alone is much larger than expected at the Sgr orbital apocenter). This star's \emph{Gaia} proper motions are $(\mu_{\rm \alpha \cos{\delta}}, \mu_\delta) = (-1.91, -0.10)\pm(0.62, 0.44)$~mas~yr$^{-1}$, and its distance error is $\sim15\%$ (the radial velocity error from APOGEE is negligible). We integrated orbits for the min/max space velocities based on the uncertainties in all measured parameters, and in neither case was the orbit bound to the Milky Way. Nonetheless, if the proper motions are systematically offset, rather than a large random error, we must await future improvements in \emph{Gaia} proper motions to address this question. Future investigation into the chemistry of HVSs across the Galaxy may shed light on the likely origin of these stars.



\section{Discussion}
\label{sec:disc}

\subsection{Implications for Chemical Tagging}

Using this \emph{k}-means clustering technique, we find that 35 of the 62 Sgr stream candidates with reliable \emph{Gaia} proper motions are likely Sgr stream members. An additional 20 stars are accreted halo stars with indistinguishable abundance patterns from the Sgr tidal debris. Through our ``weak'' chemical tagging technique, we are able to recover 55 out of 62 accreted dwarf galaxy stars in the MW halo, identified by clustering on chemical space defined by six chemical elements. Because we do not have a good idea on what the chemical abundance patterns of the Sgr core stars with [Fe/H] $<$ $-$1.2 in the APOGEE sample look like as compared to the MW, we did not extend down our analysis. However, the Sgr stream contains stars at least as metal-poor as [Fe/H] = $-$1.6 (\citealt{Chou2007,Carlin2018}), and the core contains stars as metal-poor as [Fe/H] $\sim$ $-$2.0 (\citealt{Mucciarelli2017,Hansen2018}). Future observations of the Sgr core using APOGEE-2 that target more metal-poor K giants will help demonstrate whether this technique can separate Sgr core stars from the MW stars across a much wider metallicity range to fully discover Sgr stream stars. 

\begin{figure*}[h]
\epsscale{1.0}
\plotone{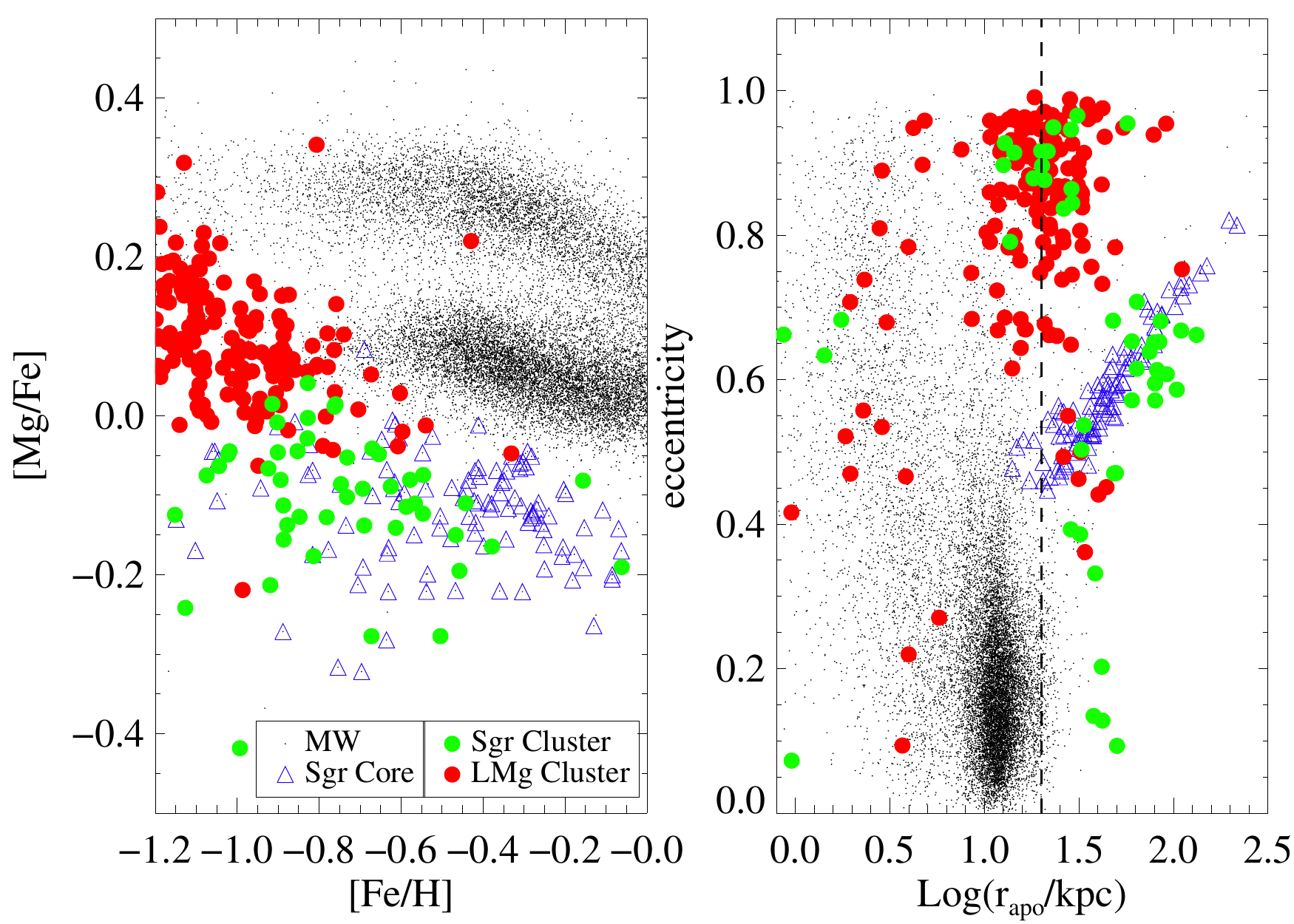} 
\caption[LMg stars]{Left: [Mg/Fe] vs. [Fe/H] abundance plane for the MW stars (black dots), Sgr core sample (purple open triangles), Sgr stream cluster (green filled-circles), and LMg cluster (red filled-circles). Right: Orbital eccentricity vs.  Log($r_{\rm apo}$) for the same samples. The dashed line signifies $r_{\rm apo}$ = 20 kpc.}
\label{fig:LMg_pop}
\end{figure*}

Thus far, we have only analyzed one cluster out of 16. While a comprehensive analysis of each individual cluster is beyond the scope of this paper, we did search the other clusters for the more metal-poor ``accreted halo'' population, identified in the APOGEE sample by \citet{Hayes2018a} and \citet{Mackereth2019}. We analyzed the [Mg/Fe] vs. [Fe/H] abundance space for each individual cluster and found that one cluster was apparently dominated by the ``Low-Mg'' population (``LMg'') identified and described by \citet{Hayes2018a} (but also described in APOGEE by \citealt{Hawkins2015} and \citealt{Fernandez-Alvar2017}) as halo stars that exhibit relatively low [Mg/Fe] abundances and halo-like kinematics. This is shown in the left panel of Figure \ref{fig:LMg_pop}.

\begin{figure*}[h]
\epsscale{1.0}
\plotone{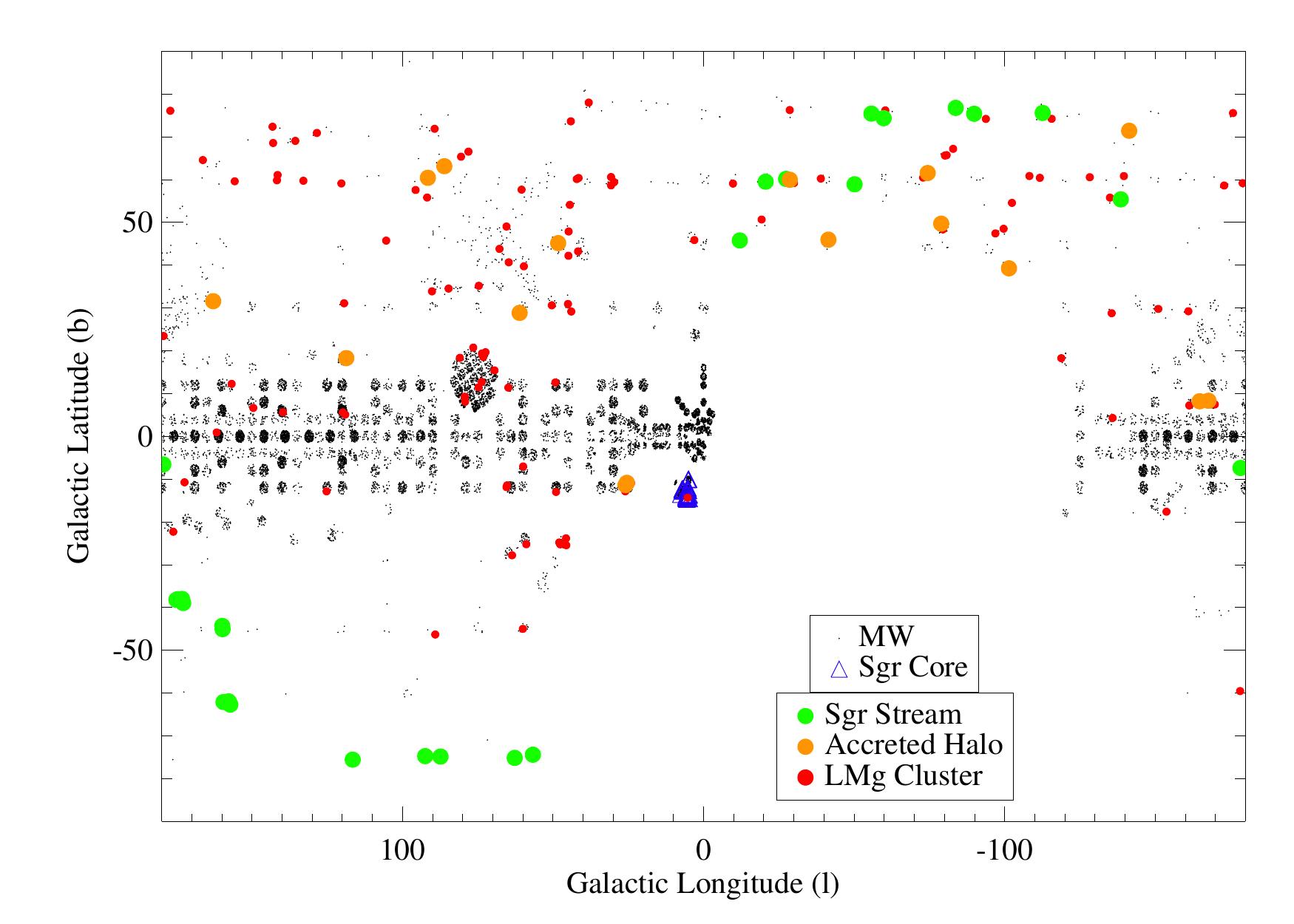} 
\caption[Map]{Galactic map of the identified stellar populations along with the MW sample on which we performed the cluster algorithm. Points are colored according to the legend.}
\label{fig:map}
\end{figure*}

\citet{Mackereth2019} was able to recover this population by conducting a \emph{k}-means clustering analysis on five-dimensional space defined by four chemical elements and orbital eccentricity. In the right panel of Figure \ref{fig:LMg_pop}, we show that the LMg population that naturally occupies one of our clusters does indeed exhibit high orbital eccentricity and median $r_{\rm apo} = $20.8 kpc, which are both characteristics of the ``accreted halo'' population thought to dominate the inner halo (\citealt{Belokurov2018,Deason2018,Helmi2018}). We show in Figure \ref{fig:map} that these stars are not in a particular region of the sky, nor are they obviously spatially clustered, as described by \citealt{Deason2018}. The recovery of this population suggests that the chemical tagging technique applied in this work is useful for identifying stars formed in relatively massive dwarf galaxies. The APOGEE IDs for the stars that fall into our LMg cluster are presented in Table \ref{tab:LMg_cand}.

\begin{table}[ht]
  \caption{LMg Cluster}
\begin{center}
\begin{tabular}{l c r}
\hline
  \textbf{APOGEE ID} & $r_{\rm apo}$ (kpc) & Ecc\\
\hline
2M00095633+6739200 & 11.59 & 0.95\\
2M00110806+8047243 & 27.65 & 0.55\\
2M00175311+6816598 & 10.71 & 0.79\\
2M01051470+4958078 & 10.66 & 0.86\\
2M02354988-0910320 & 16.98 & 0.91\\
...\tablenotemark{*} \\
\tableline
\end{tabular}
\\
\raisebox{+.4ex}{\scriptsize *}The full list is available from the online journal. \\
\end{center}
\label{tab:LMg_cand}
\end{table}




\subsection{Chemical Evolution of Sgr and the Assembly History of the MW}
\label{sec:halo}

Because we find no stream stars with [Fe/H] $>$ -0.2, which are found in abundance in the APOGEE Sgr core sample, we offer a constraint on how chemically evolved Sgr was during its most recent pericenter passage. The star formation history (SFH) of the Sgr core from \citet{Siegel2007} suggests that these stars formed $\sim$ 2 Gyr ago, which might imply that we may expect to see some of these stars stripped as we are looking at debris that \citet{Law&Majewski2010} declared was stripped as early as 1 Gyr ago. However, ff the starburst that formed this population with [Fe/H] $>$ -0.2 only occurred in the very central regions some 2 Gyr ago, then we might not see any in the stream structure.

Alternatively, \citet{Tepper-Garcia2018} argue that Sgr came into the MW gas-rich and was fully stripped of gas during its last disk passage approximately 1 Gyr ago. In this scenario, the most metal rich stars were formed during a last burst of SF coincident with this passage. The lack of stars with [Fe/H] $>$ $-$0.2 in our stream sample is also consistent with this scenario. This would require a slight tweak to the \citet{Siegel2007} SFH of Sgr, such that stars with [Fe/H] $>$ $-$0.2 formed 1 Gyr ago rather than 2-3 Gyr ago.


The ``accreted halo'' stars, which exhibit the same abundance patterns as Sgr, but are on different orbits, are likely the metal-rich end of the accreted halo population described in \citealt{Hayes2018a,Deason2018,Helmi2018,Mackereth2019}. We have shown that one of our clusters contains the more metal-poor debris. We find that these ``accreted halo'' stars, identified through chemistry alone, do indeed exhibit apocenters of $\sim$ 20 kpc along with orbital eccentricity $>$ 0.8. The chemical abundance patterns of the ``accreted halo'' stars appear to be indistinguishable from the Sgr core and stream stars at $-$1.0 $<$ [Fe/H] $<$ $-$0.6. There are potential differences in the stars with [Fe/H] $<$ $-$1.0, but we only have a small handful of Sgr stars to compare to. 

The indistinguishable chemical abundance patterns suggest that Sgr and the progenitor of the ``accreted halo'' debris exhibited very similar star formation histories. The most metal-rich of the ``accreted halo'' stars in our sample have [Fe/H] = $-$0.6. If we assume that the star formation histories must have been the same, we can use the known star formation history of Sgr to time when the progenitor of the ``accreted halo'' debris merged with the MW. Based on the star formation history of Sgr, as presented by \citet{Siegel2007}, Sgr did not form stars as metal-rich as [Fe/H] $\sim$ $-$0.6 until 8-9 Gyr ago. This is in good agreement with simulations of the ``accreted halo'' debris, which find that the progenitor was likely an LMC-sized galaxy that merged with the MW some 8-11 Gyr ago (e.g., \citealt{Belokurov2018,Mackereth2019}). 

However, from our data, we cannot determine whether this accreted halo population is from one major accretion event or a handful of smaller accretion events. While we do find stars as metal-rich as [Fe/H] $\sim$ -0.6, the mean [Fe/H] of the accreted halo population is much more metal-poor than this. If we assume that the accreted halo stars are indeed the metal-rich extension of the LMg population, then we find a mean metallicity of this population of <[Fe/H]> $\sim$ -1.0. This is an upper limit as the true mean metallicity is likely lower than this since the LMg population extends to metallicities more metal-poor than our [Fe/H] = -1.2 cutoff \citep{Hayes2018a}. Based on the stellar mass-stellar metallicity relationship described by \citet{Kirby2013}, a mean metallicity of $\lesssim$ -1.0 suggests the progenitor galaxy had a stellar mass of $\sim$ $10^{7} M_{\odot}$, similar to that of Fornax. While good arguments have been made as to why this progenitor galaxy need  be massive (e.g., \citealt{Belokurov2018,Mackereth2019}), such a galaxy would fall considerably off the stellar-mass stellar-metallicity relationship, being too metal-poor for its mass.

\section{Conclusions}

We have demonstrated a ``weak chemical tagging'' method that efficiently identifies Sgr stream stars from chemical abundances alone. Specifically, we performed a \emph{k}-means clustering analysis on six APOGEE chemical abundance ratios in which Sgr core stars are distinct from Milky Way stellar populations \citep{Hasselquist2017}. As a test, our technique recovers 19 of 21 kinematically-selected Sgr stream members. Of the 62 stars in our chemically-selected APOGEE sample that have reliable \textit{Gaia} proper motions, 35 ($56\%$) are confirmed to be Sgr stream stars, and 20 more belong to an ``accreted halo'' population that apparently spawned from a progenitor with a very similar SFH to Sgr in its first 2-5 Gyr of existence. Therefore, our detection reliability of ``Sgr-like'' galaxy stars is $\sim89\%$. 


To explore the nature of the chemically selected candidates, we integrated orbits using \textit{Gaia} proper motions and APOGEE radial velocities and distance estimates. A total of 35 candidates have positions and 3D velocities consistent with the \citet{Law&Majewski2010} model of Sgr disruption. The majority of these correspond to Sgr tidal debris from the last two pericenter passages of Sgr's orbit. An additional 20 stars are on highly eccentric orbits (median $e = 0.9$) with median $r_{\rm apo}= 20.2$ kpc. We refer to this sample as the ``accreted halo'' stars, because their kinematics appear similar to those of the stars recently claimed to originate in an ancient major merger (e.g., \citealt{Schuster2012,Belokurov2018, Deason2018}). These stars are mostly not located near the Sgr debris plane, but because they were selected to have chemistry similar to Sgr, it is possible that they are related to the oldest debris from the Sgr disruption. It may also be that they were contributed by a different merger event whose progenitor was similar to Sgr. 

Also included in the Sgr chemical cluster are 4 apparent bulge stars. These stars were included in the Sgr cluster due to their exceptionally low [O/Fe] abundance, but exhibited slightly enhanced [Ni/Fe] abundance, uncharacteristic of Sgr stars. However, they are outliers in chemical space, as compared to the bulk of the bulge stars in APOGEE, and suggest that additional chemical information is required to separate chemically similar (but not chemically identical) groups. We additionally find one star whose 3D position in the Galaxy is consistent with being part of the Sgr debris at the most recent apocenter, but whose total Galactocentric velocity is $\sim600$~km~s$^{-1}$. The orbit of this star does not approach within 50 kpc of the Galactic center, making it inconsistent with most scenarios for producing hypervelocity stars. Because it has abundances consistent with Sgr, it is possible that this star was somehow ejected from the Sgr stream, but further refinement of the proper motion may be necessary to understand its origin. 

Future APOGEE-2 observations of the Sgr core that extend to lower metallicities will allow us to test this ``tuned'' \emph{k}-means clustering algorithm to see if we can recover more metal-poor Sgr stars that we know to be in the streams. This will also allow for a more in-depth analysis of the similarities between the ``accreted halo'' population of stars thought to be from a massive progenitor from 8+ Gyr ago and the Sgr stars. 


\vspace{0.5cm}
\scriptsize{\emph{Acknowledgements.} Funding for the Sloan Digital Sky Survey IV has been provided by the
Alfred P. Sloan Foundation, the U.S. Department of Energy Office of
Science, and the Participating Institutions. SDSS acknowledges
support and resources from the Center for High-Performance Computing at
the University of Utah. The SDSS web site is www.sdss.org.

SDSS is managed by the Astrophysical Research Consortium for the Participating Institutions of the SDSS Collaboration including the Brazilian Participation Group, the Carnegie Institution for Science, Carnegie Mellon University, the Chilean Participation Group, the French Participation Group, Harvard-Smithsonian Center for Astrophysics, Instituto de Astrof{\'{\i}}sica de Canarias, The Johns Hopkins University, Kavli Institute for the Physics and Mathematics of the Universe (IPMU) / University of Tokyo, Lawrence Berkeley National Laboratory, Leibniz Institut f{\"u}r Astrophysik Potsdam (AIP), Max-Planck-Institut f{\"u}r Astronomie (MPIA Heidelberg), Max-Planck-Institut f{\"u}r Astrophysik (MPA Garching), Max-Planck-Institut f{\"u}r Extraterrestrische Physik (MPE), National Astronomical Observatory of China, New Mexico State University, New York University, University of Notre Dame, Observat{\'o}rio Nacional / MCTI, The Ohio State University, Pennsylvania State University, Shanghai Astronomical Observatory, United Kingdom Participation Group, Universidad Nacional Aut{\'o}noma de M{\'e}xico, University of Arizona, University of Colorado Boulder, University of Oxford, University of Portsmouth, University of Utah, University of Virginia, University of Washington, University of Wisconsin, Vanderbilt University, and Yale University.

SH is supported by an NSF Astronomy and Astrophysics Postdoctoral Fellowship under award AST-1801940. AGH and OZ acknowledge support provided by the Spanish Ministry of Economy and Competitiveness (MINECO) under grant AYA-2017-88254-P. We thank Adrian Price-Whelan for assistance with running and modifying the \texttt{gala} software.


\normalsize

\bibliographystyle{apj}
\bibliography{/home/users/sten/Papers/ref_og.bib}
 
\end{document}